\documentclass[twocolumn]{aastex62}
\usepackage{natbib}

\newcommand{\Mh}{M_{\rm h}}
\newcommand\Msun{M_{\odot}}
\newcommand\Mmin{M_{\min}}
\newcommand\Mstar{M_{\star}}

\newcommand{\Mpivot}{M_{\rm h}^{\rm pivot}}
\newcommand\zbest{z_{p, {\rm best}}}
\newcommand\zp{z_{p}}
\newcommand\fsat{f_{\rm s}}
\newcommand\bg{b_{\rm g}}
\newcommand\bh{b_{\rm h}}
\newcommand\sigmaM{\sigma_{\log{M}}}
\newcommand\logMstarlimit{\log_{10}(M_{\star, {\rm limit}}/h^{-2}\Msun)}
\newcommand\logMstarh{\log_{10}(M_{\star}/h^{-2}\Msun)}

\newcommand\logSFR{\log_{10}({\rm SFR}/h^{-2}\Msun {\rm yr}^{-1})}
\newcommand\logSFRlimit{\log_{10}({\rm SFR_{limit}}/h^{-2}\Msun {\rm yr}^{-1})}

\shorttitle{The Subaru HSC Photo-$z$ Galaxy Clustering I}
\shortauthors{Ishikawa et al.}

\begin{document}

\title{The Subaru HSC Galaxy Clustering with Photometric Redshift I: Dark Halo Masses Versus Baryonic Properties of Galaxies at $0.3 \leq z \leq 1.4$}

\correspondingauthor{Shogo Ishikawa}
\email{shogo.ishikawa@nao.ac.jp}

\author{Shogo Ishikawa}
\affil{Center for Computational Astrophysics, National Astronomical Observatory of Japan, Mitaka, Tokyo 181-8588, Japan}
\affil{National Astronomical Observatory of Japan, Mitaka, Tokyo 181-8588, Japan}

\author{Nobunari Kashikawa}
\affil{Department of Astronomy, School of Science, The University of Tokyo, Tokyo 113-0033, Japan}

\author{Masayuki Tanaka}
\affil{National Astronomical Observatory of Japan, Mitaka, Tokyo 181-8588, Japan}
\affil{Department of Astronomical Science, SOKENDAI (Graduate University for Advanced Studies), Mitaka, Tokyo 181-8588, Japan}

\author{Jean Coupon}
\affil{Astronomy Department, University of Geneva, Chemin d'Ecogia 16, CH-1290 Versoix, Switzerland}

\author{Alexie Leauthaud}
\affil{Department of Astronomy and Astrophysics, University of California, Santa Cruz, 1156 High Street, Santa Cruz, CA 95064, US}

\author{Jun Toshikawa}
\affil{Institute for Cosmic Ray Research, The University of Tokyo, Kashiwa, Chiba 277-8582, Japan}
\affil{Department of Physics, University of Bath, Claverton Down, Bath, BA2 7AY, UK}

\author[0000-0002-4377-903X]{Kohei Ichikawa}
\affil{Astronomical Institute, Graduate School of Science, Tohoku University, 6-3 Aramaki, Aoba-ku, Sendai 980-8578, Japan}
\affil{Frontier Research Institute for Interdisciplinary Sciences, Tohoku University, Sendai 980-8578, Japan}

\author{Taira Oogi}
\affil{Kavli Institute for the Physics and Mathematics of the Universe (WPI), The University of Tokyo Institutes for Advanced Study, The University of Tokyo, 5-1-5 Kashiwanoha, Kashiwa, Chiba 277-8583, Japan}

\author{Hisakazu Uchiyama}
\affil{National Astronomical Observatory of Japan, Mitaka, Tokyo 181-8588, Japan}

\author{Yuu Niino}
\affil{Research Center for the Early Universe, The University of Tokyo, 7-3-1 Hongo, Bunkyo, Tokyo 113-0033, Japan}
\affil{ Institute of Astronomy, University of Tokyo, 2-21-1 Osawa, Mitaka, Tokyo 181-0015, Japan}

\author{Atsushi J. Nishizawa}
\affil{Institute for Advanced Research, Nagoya University, Furo-cho, Chikusa-ku, Nagoya 464-8602, Japan}

\begin{abstract}
We present the clustering properties of low-$z$ $(z\leq1.4)$ galaxies selected by the Hyper Suprime-Cam Subaru Strategic Program Wide layer over $145$ deg$^{2}$. 
The wide-field and multi-wavelength observation yields $5,064,770$ galaxies at $0.3\leq z\leq1.4$ with photometric redshifts and physical properties. 
This enables the accurate measurement of angular correlation functions and subsequent halo occupation distribution (HOD) analysis allows the connection between baryonic properties and dark halo properties. 
The fraction of less-massive satellite galaxies at $z\lesssim1$ is found to be almost constant at $\sim20\%$, but it gradually decreases beyond $\Mstar \sim 10^{10.4}h^{-2}\Msun$. 
However, the abundance of satellite galaxies at $z>1$ is quite small even for less-massive galaxies due to the rarity of massive centrals at high-$z$. 
This decreasing trend is connected to the small satellite fraction of Lyman break galaxies at $z>3$. 
The stellar-to-halo mass ratios at $0.3\leq z\leq1.4$ are almost consistent with the predictions obtained using the latest empirical model; however, we identify small excesses from the theoretical model at the massive end. 
The pivot halo mass is found to be unchanged at $10^{11.9-12.1}h^{-1}\Msun$ at $0.3\leq z\leq1.4$, and we systematically show that $10^{12}h^{-1}\Msun$ is a universal pivot halo mass up to $z\sim5$ that is derived using only the clustering/HOD analyses. 
Nevertheless, halo masses with peaked instantaneous baryon conversion efficiencies are much smaller than the pivot halo mass regardless of a redshift, and the most efficient stellar-mass assembly is thought to be in progress in $10^{11.0-11.5}h^{-1}\Msun$ dark haloes.
\end{abstract}


\section{Introduction} \label{sec:intro}
The $\Lambda$CDM paradigm predicts that small density fluctuations in the early Universe that are imprinted in the cosmic microwave background evolves into the large-scale structure of the Universe \citep[e.g.,][]{davis85,jenkins98,springel05}. 
In the context of the $\Lambda$CDM cosmological model, dark matter forms self-bounding clumps, known as dark haloes, and galaxies are born in the center of these haloes by trapping baryonic matters using their gravitational potential \citep[e.g.,][]{white78,blumenthal84,white91,cole94,mo96}. 
Therefore, galaxies are biased tracers towards invisible underlying dark matter distribution. 
Many studies have attempted to elucidate the biased relationship between dark matter and baryons, based on the notable success in mapping the 3D distribution of galaxies using large extensive galaxy redshift surveys \citep[e.g.,][]{delapparent86,peacock01,zehavi05}. 

To achieve an in-depth understanding of the co-evolution between visible baryonic matter and invisible dark matter, some observational methodologies for the measurement of dark halo mass, which is one of the fundamental parameters for galaxy-formation efficiency, have been proposed and performed over several decades including X-ray observations \citep[e.g.,][]{evrard96,vikhlinin06,piffaretti08,kravtsov18}, measurements of a weak-lensing signal \citep[e.g.,][]{smith01,mandelbaum06} and kinematics of satellite galaxies \citep[e.g.,][]{more11, wojtak13}, and the use of two-point statistics of galaxy clustering \citep[e.g.,][]{peebles80,zehavi11}. 
However, almost all observational approaches have their pros and cons. 
For example, X-ray observations provide information on the total gravitational masses of individual groups/clusters of galaxies by assuming the hydrostatic equilibrium state for their intra-cluster medium. 
However, this method can only be applied to massive objects in the local Universe ($\Mh \gtrsim 10^{14}\Msun$ haloes at $z\lesssim0.2$ in typical cases). 
Satellite kinematics can directly connect the baryonic properties of central galaxies to their host dark halo mass by measuring the velocity dispersion of orbiting satellite galaxies. 
Weak-lensing signals are indicative of foreground matter-density fluctuations based on observations of distorted shapes of background galaxies. 
However, these techniques can only be used for low-$z$ galaxies (only for $z<1$ galaxies in typical cases) due to the faintness of key signals. 

Compared to these observational approaches, galaxy clustering signal provides averaged statistical information on galaxy distribution and underlying matter fluctuation of focusing galaxies. 
Galaxy auto-correlation functions indicate how strong galaxies are bound to each other by their gravity. 
However, they can be easily measured by assessing the gravitational interaction between galaxies. 
Moreover, constraining galaxy bias via galaxy clustering observations has been succeessful in extending the redshift range compared to other galaxy--halo connection techniques \citep[][for a review of the galaxy--halo connection]{wechsler18}, even at $z>7$ \citep{barone-nugent14}. 
In addition, halo-model approaches, such as the abundance-matching (AM) technique \citep[e.g.,][]{kravtsov99,vale04,conroy06} and the halo occupation distribution (HOD) model \citep[e.g.,][]{seljak00,ma00,berlind02,berlind03,vdbosch03}, facilitate the derivation of the properties of dark haloes by assuming that dark haloes hosting galaxies are virialized, and characterizing the level of galaxy bias by defining the conditional probability as $P(N|\Mh)$ \citep[][for a review of halo-model approach]{cooray02}. 
Furthermore, recent massively parallel large hydrodynamical simulations such as the Illustris/IllustrisTNG simulations \citep{vogelsberger14,springel18} and the EAGLE simulations \citep{schaye15,crain15} have contributed to our understanding and interpretation of galaxy clustering by direct comparison between the simulated visible and invisible Universe. 

Towards constraining the biased relationship between dark matter and baryons, and elucidating the physical processes of galaxy formation and evolution, auto-correlation functions of various galaxy populations have been measured by utilizing the cutting-edge observational data of spectroscopic galaxy redshift surveys (e.g., SDSS, BOSS, VIPERS, and FastSound), as well as extensive multi-wavelength photometric surveys (e.g., COSMOS, UltraVISTA, CFHTLS, and DES). 
Given that accurate positions on three-dimensional coordinate are available, spectroscopic surveys have significant advantages in clustering studies; precise two-dimensional auto-correlation functions can be evaluated by projecting real-space correlation functions. 
This real-space correlation function contains cosmological information and can be used to test the general relativistic theory and constrain the cosmic growth rate by analyzing the redshift-space distortion effect \citep[e.g.,][]{raccanelli13,alam17}. 

However, clustering studies that use extensive photometric surveys have unique characteristics and advantages. 
First, photometric surveys can be used to select faint galaxies. 
Galaxy redshift surveys are confined to bright objects in order to detect apparent emission line(s), whilst photometric surveys can be used to select faint galaxies even if they only satisfy the detection limit of each image. 
Therefore, compared to spectroscopic surveys, galaxy clustering studies based on photometric surveys can address a variety of galaxy populations and baryonic characteristics. 
Moreover, the targeted redshift range is further broadened for studies based on photometric surveys; i.e., clustering-signal measurement using spectroscopic observations is up to $z\sim3$ \citep{durkalec15,durkalec18} and typically confined to $z<1$ due to the strict flux threshold, whereas those based on photometric data exceed $z\sim5$ with a high S/N ratio \citep[e.g.,][]{hildebrandt09,ishikawa17,harikane18}. 
Second, photometric observations are free from the fiber/slit-collision effect. 
In fiber spectroscopic observations, it is necessary to allocate fibers on a fiber plug plate. 
Moreover, the minimum physical scale between galaxy pairs is limited by the finite physical size of fibers. 
Similar limitation for slit allocation is inevitable in slit spectroscopy. 
A minimum physical scale in photometric observations is determined by a seeing size and small-scale clustering, which is a key clustering signal to constrain the satellite galaxy formation, can be well measured. 

In this paper, we aim to describe the relationship between galaxy clustering properties and the fundamental physical parameters of baryons; i.e., galaxy stellar mass and star-formation rate (SFR), with high-precision statistics based on large galaxy samples obtained by the Hyper Suprime-Cam Subaru Strategic Program \citep[HSC SSP;][]{aihara18a} Wide layer S16A internal Data Release \citep{aihara18b}. 
The unique capability of the Subaru Hyper Suprime-Cam \citep[HSC;][]{miyazaki18} and the latest sophisticated SED-fitting technique \citep{tanaka15} facilitates the acquisition unbiased $\sim5,000,000$ photometric-redshift selected galaxies over $\sim145$ deg$^{2}$ with deep limiting magnitude down to $i=25.9$ compared to other wide-field surveys. 
These characteristics of the HSC SSP Wide layer allow the selection of a sufficiently large number of both less-massive galaxies ($\Mstar\sim10^{9}\Msun$), as well as massive galaxies ($\Mstar\sim10^{11}\Msun$) over a wide redshift range ($0.3\leq z\leq1.4$), to precisely study the redshift evolution and physical parameter dependence of galaxy clustering. 

This is the first paper on a series of clustering analyses using the photometric-redshift catalogue of the HSC SSP. 
We will extend the targeted redshift range and stellar-mass limit using photometric catalogues in the Deep/UltraDeep layers in addition to galaxy populations, by dividing them into star-forming galaxies and passive galaxies, and investigate the evolutionary history of galaxies by connecting the results presented in this paper to future studies. 
The main goal of this series of study on photo-$z$ galaxy clustering is to understand and gain insights into the galaxy--dark halo co-evolution history at $z\lesssim3$ using common galaxy catalogues, and the uniform analysis method. 

This paper is organized into several sections. 
In Section~2, we describe the details of the photometric data and the sample selection method using the photometric-redshift catalogue generated by the HSC SSP S16A data. 
In Section~3, we review the methodology of the clustering and HOD analysis, and the auto-correlation functions of our galaxy samples obtained via clustering analysis and the results of the HOD-model analysis are shown in Section~4. 
We discuss the relationship between galaxies and their host dark haloes at $0.3\leq z \leq 1.4$, and compare them with other observational studies as well as numerical simulations and theoretical models in Section~5. 
A summary of major findings based on halo-model analyses are presented in Section~6. 

All of the photometric magnitudes are quoted in the Absolute Bolometric (AB) magnitude system \citep{oke83}. 
Throughout this paper, we employ the {\it Planck} 2015 cosmological parameters: the density parameters are $\Omega_{\rm m} = 0.309$, $\Omega_{\Lambda} = 0.691$, and $\Omega_{\rm b} = 0.049$, the Hubble parameter is assumed as $H_{0} = 100h$ with the dimensionless Hubble parameter as $h = 0.677$, the normalization of the matter fluctuation power spectrum is $\sigma_{8} = 0.816$, and the spectral index of the primordial power spectrum is $n_{\rm s} = 0.967$, respectively \citep{planck16}. 
Galaxy stellar mass and dark halo mass are represented as $\Mstar$ and $\Mh$, respectively. 
All of logarithm in this paper are common logarithm with base $10$. 

\section{Data and Sample Selection} \label{sec:data}
\subsection{Photometric Data} \label{subsec:photodata}
We use the HSC SSP S16A internal data that were released in August 2016, and obtained between March 2014 to April 2016 \citep{aihara18b}. 
The HSC SSP data is composed of three layers: Wide, Deep, and the UltraDeep layer \citep{aihara18a}. 
In this study, photometric data is obtained only in the Wide layer with the aim of achieving a high S/N ratio of clustering signals and imposing strong constraints on massive ends at $z\leq1.4$. 
The data of the Deep and UltraDeep layers will be used in our future studies to connect faint, high-redshift galaxies to their host dark haloes using clustering/HOD analyses. 

The HSC SSP Wide layer is composed of six distinct patches: XMM-LSS, GAMA09H, GAMA15H, HectoMAP, VVDS, and WIDE12H. 
The objective of the Wide layer is to observe $1,400$ deg$^{2}$ in total by $916$ pointings with $1.8$ deg$^{2}$ field-of-views of the HSC \citep{miyazaki18} when the HSC SSP is completed. 
However, the data of the S16A is available over $\sim 178$ deg$^{2}$. 
Each patch of the Wide layer is observed using five optical wavelengths: $g$-, $r$-, $i$-, $z$-, and $y$-band \citep{kawanomoto18} that satisfy depths of $g<26.5$, $r<26.1$, $i<25.9$, $z<25.1$, and $y<24.4$ ($5\sigma$ limiting magnitudes with $2$ arcsec diameter apertures), respectively. 

Data of the HSC SSP are reduced using the optical imaging data processing pipeline, {\tt hscpipe} \citep
{bosch18}. 
The {\tt hscpipe} is developed based on the pipeline of the Large Synoptic Survey Telescope \citep[LSST; ][]{ivezic08,axelrod10,juric15}. 
Pixels around bright stars are masked according to the procedure of \citet{coupon18}. 
We also exclude pixels at the edge of photometric images, on cosmic rays, saturated pixels, and pixels with flags of {\tt flags\_pixel\_edge}, {\tt flags\_pixel\_interpolated\_center}, {\tt flags\_pixel\_saturated\_center}, {\tt flags\_pixel\_cr\_center}, and {\tt flags\_pixel\_bad}  \citep{aihara18b}. 
The total survey area of this paper is $144.68$ deg$^{2}$ after masking. 

\subsection{Photometric-redshift Catalogue} \label{subsec:photo-z}
The HSC SSP provides multiple photometric-redshift catalogues \citep{tanaka18}. 
In this study, we use an HSC photometric-redshift catalogue constructed by using an SED-fitting code with Beysian physical priors, {\sc Mizuki} \citep{tanaka15}. 
We briefly describe characteristics of the SED-fitting code below. 
See \citet{tanaka15} for more details. 

{\sc Mizuki} is a template-fitting code and galaxy SED templates are generated by {\sc GALAXEV}, which is the spectral synthesis model developed by \citet{bruzual03}. 
A star-formation history of galaxy SED templates is assumed an exponential-time decay model with varying the declination time scale, $\tau$. 
The solar metallicity abundance is only assumed for the SED templates; however, results of the SED fitting are not significantly changed when including the SED templates with sub-solar metallicity abundance. 
The initial-mass function (IMF) is assumed a Chabrier IMF \citep{chabrier03} and the dust attenuation follows the Calzetti curve with varying the optical depth, $\tau_{\rm V}$ \citep{calzetti00}. 
Nebular emission lines are added to the galaxy SED templates generated by \citet{bruzual03}, with the intensity ratios of \citet{inoue11} and the differential dust extinction law proposed by \citet{calzetti97}. 

The best-fitting photometric redshifts are evaluated through the likelihood: 
\begin{equation}
{\mathcal L} \propto \exp{\left(- \chi^{2}_{\rm SED fit}/2 \right)}, 
\label{eq:likelihood}
\end{equation}
where the $\chi^{2}_{\rm SED fit}$ is a value of the chi-square fitting of the SED-fitting procedure and can be computed as follows: 
\begin{equation}
\chi^{2}_{\rm SED fit} = \sum_{i} \frac{\left( f_{i, {\rm obs}} - \alpha f_{i, {\rm model}} \right)^{2}}{\sigma_{i, {\rm obs}}^{2}}. 
\label{eq:chisq}
\end{equation}
Here, $f_{i, {\rm obs}}$ and $f_{i, {\rm model}}$ are the observed and model SED fluxes of the $i$th filter, $\sigma_{i, {\rm obs}}$ is the uncertainty of the $i$th observed flux, and $\alpha$ is a normalization parameter that controls the amplitude of the model SED. 
Physical priors are multiplied by the likelihood to obtain posteriors. 

The HSC photometric-redshift catalogue using {\sc Mizuki} is constructed by independently detecting sources from HSC science image frames of each optical band ($g$-, $r$-, $i$-, $z$-, and $y$-band). 
The detection threshold is $5\sigma$, where $\sigma$ is a dispersion of the sky fluctuation, compared to RMSs by convolving the point-spread function. 
The fluxes of detected objects are measured without matching their shapes and centroids across optical bands to determine the reference band for each object, then the cModel fluxes of each objects are measured with matching shapes and centroids to their values on reference images ($i$-band images for most case).
 
The SED-fitting technique is applied to objects with clean cModel magnitudes at least three bands. 
In the construction of galaxy samples for this study, we set the magnitude limits as $g\leq26.0$, $r\leq25.6$, $i\leq25.4$, $z\leq24.2$, and $y\leq23.4$, and the brightest magnitude of each band is $18.0$ for the selection of reliable galaxy samples. 
Refer to \citet{aihara18b,tanaka18,bosch18} for more information. 

\subsection{Galaxy Sample Selection and Physical Properties} \label{subsec:sample_selection}
Galaxy samples are extracted from the HSC photometric-redshift catalogue using {\sc Mizuki}. 
We use ``photo-$z$ best ($\zbest$)'' parameter, which is evaluated to minimize a risk parameter in the SED fitting procedure, for characterizing the redshift of each object. 
To avoid selecting galaxies with a multiple-peak PDF or a catastrophic photo-$z$ error, we set conditions of $\chi^{2}_{{\rm SEDfit}} / {\rm d.o.f.} \leq 3.0$ and ${\rm risk} \leq 0.1$. 
Refer to \citet{tanaka18} for more details of photometric-redshift catalogues of the HSC SSP. 
Hereafter, we quote the photo-$z$ best parameter as a photometric redshift of each galaxy. 

\begin{figure}[tbp]
\epsscale{1.25}
\plotone{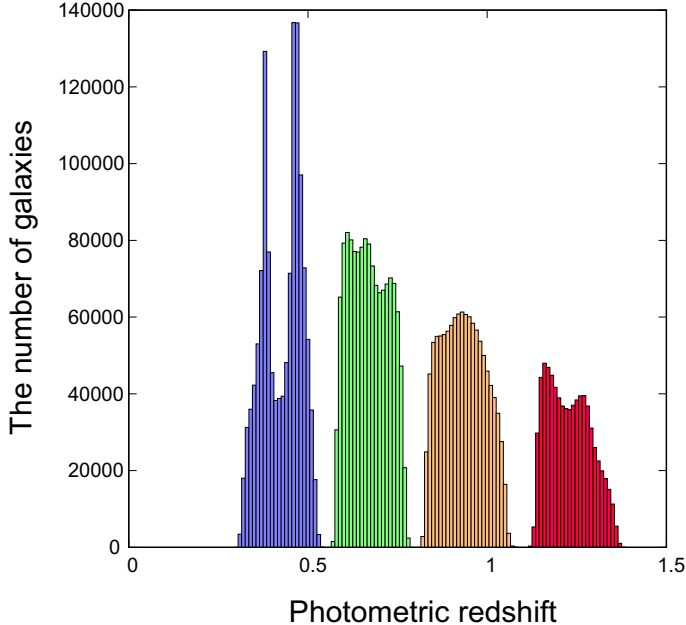}
\caption{The redshift distribution of our HSC photometric-redshift selected galaxy samples. We adopt the $\zbest$ parameter as the photometric redshift of each galaxy. To trace the redshift evolution of galaxy clustering properties, galaxy samples are divided into four distinct redshift bins: $0.30 \leq \zbest < 0.55$ ($z_{1}$; blue), $0.55 \leq \zbest < 0.80$ ($z_{2}$; green), $0.80 \leq \zbest < 1.10$ ($z_{3}$; orange), and $1.10 \leq \zbest \leq 1.40$ ($z_{4}$; red). }
\label{fig:Nz}
\end{figure}
We divided our photo-$z$ galaxy samples into four redshift bins: $0.30 \leq z_{1} < 0.55$, $0.55 \leq z_{2} < 0.80$, $0.80 \leq z_{3} < 1.10$, and $1.10 \leq z_{4} \leq 1.40$, to investigate the redshift evolution of the clustering properties. 
It is noted that our targeted redshift starts from $z=0.30$ in order to capture the Balmer/4000${\rm \AA}$ break using $g$- and $r$-band photometry. 
The total survey comoving volume of this study is $\sim 1.13$ Gpc$^{3}$ and the total number of galaxies that satisfy the aforementioned condition is $5,064,770$. 

Galaxies with large photo-$z$ uncertainties can contaminate other redshift bins (foreground/background galaxies account for $\sim 15\%$ ($30\%$) of the total galaxies at $0.4<z<0.6$ ($1.2<z<1.4$) for this photometric-redshift galaxy catalogue). 
To construct a pure redshift-binned galaxy catalogue by avoiding redshift bin-to-bin contaminations, we select galaxies that satisfy the following condition: 
\begin{equation}
(z_{i, {\rm min}} \leq \zbest - dz_{-1\sigma}) \cap (\zbest + dz_{+1\sigma} \leq z_{i, {\rm max}}), 
\label{eq:z_condition}
\end{equation}
where $dz_{+1\sigma}$ $(dz_{-1\sigma}$) represents the upper (lower) bound of the $1\sigma$ confidence interval and $z_{i, {\rm max}}$ ($z_{i, {\rm min}}$) is the upper (lower) bound of $i$th redshift bin. 
The redshift distribution of our selected galaxies is shown in Figure~\ref{fig:Nz}. 
The number of galaxies in the final catalogue of this study is $4,531,285$. 

In Figure~\ref{fig:Nz}, it is observed that the $z_{1}$ bin has two peaks in the redshift distribution. 
This may be due to the Balmer/4000${\rm \AA}$ break, which is a key spectral feature to determine photometric redshift, falls into the gap between $g$- and $r$-band transmission wavelength and the weak constraint on $\zp \lesssim 0.5$ caused by the lack of NUV photometry. 
However, this double-peak feature in the redshift distribution is not problematic in this study because galaxy clustering signals are measured by projecting the galaxies along the line-of-sight for each redshift bin and almost all the galaxies in the $z_{1}$ bin identify the same redshift bin within $1\sigma$ uncertainty. 

\begin{table*}[tbp]
\begin{center}
\caption{Details of Cumulative Stellar-mass Limited Subsamples}
\begin{tabular}{lccccccccccc} \hline \hline
 & \multicolumn{2}{c}{$z_{1}$} & & \multicolumn{2}{c}{$z_{2}$} & & \multicolumn{2}{c}{$z_{3}$} & & \multicolumn{2}{c}{$z_{4}$}  \\ 
& \multicolumn{2}{c}{$0.30 \leq z < 0.55$} & & \multicolumn{2}{c}{$0.55 \leq z < 0.80$}  & & \multicolumn{2}{c}{$0.80 \leq z < 1.10$} & & \multicolumn{2}{c}{$1.10 \leq z \leq 1.40$}\\ 
\cline{2-3}\cline{5-6}\cline{8-9}\cline{11-12}
Stellar-mass limit\footnote{Threshold stellar mass of each subsample in units of $h^{-2}\Msun$ in a logarithmic scale.} & \multicolumn{1}{c}{$N$\footnote{The number of galaxies of each subsample. }} & \multicolumn{1}{c}{$M_{\star, {\rm med}}$\footnote{Median stellar mass of each subsample in units of $h^{-2}\Msun$ in a logarithmic scale. }} & & \multicolumn{1}{c}{$N$} & \multicolumn{1}{c}{$M_{\star, {\rm med}}$} & & \multicolumn{1}{c}{$N$} & \multicolumn{1}{c}{$M_{\star, {\rm med}}$} & & \multicolumn{1}{c}{$N$} & \multicolumn{1}{c}{$M_{\star, {\rm med}}$} \\ \hline
$8.60$ & $892,206$ & $9.27$ & & -- & -- & & -- & -- & & -- & -- \\
$8.80$ & $727,385$ & $9.45$ & & -- & -- & & -- & -- & & -- & -- \\
$9.00$ & $592,648$ & $9.62$ & & $1,038,903$ & $9.72$ & & -- & -- & & -- & -- \\
$9.20$ & $479,783$ & $9.79$ & & $876,600$ & $9.86$ & & -- & -- & & -- & -- \\
$9.40$ & $384,732$ & $9.94$ & & $723,725$ & $10.00$ & & 818,934 & $9.85$ & & -- & -- \\
$9.60$ & $305,332$ &$10.08$  & & $590,583$ & $10.12$ & & $625,246$ & $10.00$ & & -- & -- \\
$9.80$ & $235,427$ & $10.21$ & & $470,783$ & $10.24$ & & $447,467$ & $10.18$ & & $461,539$ & $10.11$ \\
$10.00$ & $173,907$ & $10.35$ & & $361,233$ & $10.37$ & & $314,364$ & $10.36$ & & $306,547$ & $10.26$ \\
$10.20$ & $121,214$ & $10.48$ & & $257,323$ & $10.50$ & & $216,952$ & $10.53$ & & $181,690$ & $10.43$ \\
$10.40$ & $76,394$ & $10.62$ & & $166,123$ & $10.64$ & & $145,143$ & $10.67$ & & $101,175$ & $10.61$ \\
$10.60$ & $41,422$ & $10.77$ & & $93,603$ & $10.78$ & & $88,953$ & $10.81$ & & $52,664$ & $10.80$ \\
$10.80$ & $17,827$ & $10.92$ & & $43,658$ & $10.94$ & & $46,228$ & $10.96$ & & $26,123$ & $10.98$  \\
$11.00$ & $4,967$ & $11.08$ & & $15,944$ & $11.11$ & & $18,878$ & $11.12$ & & $12,222$ & $11.16$ \\
$11.20$ & -- & -- & & $4,356$ & $11.30$ & & $5,662$ & $11.29$ & & $4,960$ & $11.31$ \\  \hline\hline
\label{tab:zbin}
\end{tabular}
\end{center}
\end{table*}

\begin{table*}[tbp]
\begin{center}
\caption{Details of Cumulative SFR-limited Subsamples}
\begin{tabular}{lccccccccccc} \hline \hline
 & \multicolumn{2}{c}{$z_{1}$} & & \multicolumn{2}{c}{$z_{2}$} & & \multicolumn{2}{c}{$z_{3}$} & & \multicolumn{2}{c}{$z_{4}$}  \\ 
& \multicolumn{2}{c}{$0.30 \leq z < 0.55$} & & \multicolumn{2}{c}{$0.55 \leq z < 0.80$}  & & \multicolumn{2}{c}{$0.80 \leq z < 1.10$} & & \multicolumn{2}{c}{$1.10 \leq z \leq 1.40$}\\ 
\cline{2-3}\cline{5-6}\cline{8-9}\cline{11-12}
SFR limit\footnote{Threshold SFR of each subsample in units of $h^{-2}\Msun {\rm yr}^{-1}$ in a logarithmic scale. } & \multicolumn{1}{c}{$N$\footnote{The number of galaxies of each subsample. }} & \multicolumn{1}{c}{${\rm SFR}_{\rm med}$\footnote{Median SFR of each subsample in units of $h^{-2}\Msun {\rm yr}^{-1}$ in a logarithmic scale. }} & & \multicolumn{1}{c}{$N$} & \multicolumn{1}{c}{${\rm SFR}_{\rm med}$} & & \multicolumn{1}{c}{$N$} & \multicolumn{1}{c}{${\rm SFR}_{\rm med}$} & & \multicolumn{1}{c}{$N$} & \multicolumn{1}{c}{${\rm SFR}_{\rm med}$} \\ \hline
$-1.50$ & $728,514$ & $-0.40$ & & $879,443$ & $-0.14$ & & -- & -- & & -- & -- \\
$-1.00$ & $687,106$ & $-0.37$ & & $824,188$ & $-0.11$ & & $777,669$ & $0.12$ & & -- & -- \\
$-0.50$ & $432,225$ & $-0.15$ & & $685,552$ & $-0.01$ & & $671,234$ & $0.19$ & & $419,436$ & $0.67$ \\
$0.00$ & $150,447$ & $0.23$ & & $333,221$ & $0.28$ & & $479,151$ & $0.33$ & & $370,153$ & $0.72$ \\
$0.50$ & $25,761$ & $0.65$ & & $79,414$ & $0.67$ & & $141,518$ & $0.69$ & & $280,701$ & $0.83$\\
$1.00$ & -- & -- & & -- & -- & & $15,806$ & $1.12$ & & $83,908$ & $1.18$ \\ \hline\hline
\label{tab:zbin_sfr}
\end{tabular}
\end{center}
All galaxies of SFR sample satisfy the stellar-mass limit of each redshift bin. 
\end{table*}

Apart from the photometric redshifts, we also use galaxy stellar masses and SFRs calculated by via SED fitting to characterize the baryonic properties of galaxies. 
Stellar masses and SFRs are median values, $M_{\star, {\rm med}}$ and ${\rm SFR}_{\rm med}$, that are evaluated by integrating the PDFs of each parameter: 
\begin{equation} 
\int_{M_{\star, {\rm min}}}^{M_{\star, {\rm med}}} P(\Mstar) d\Mstar = 0.5, 
\label{eq:Mstar_med}
\end{equation}
and 
\begin{equation}
\int_{{\rm SFR}_{\rm min}}^{{\rm SFR}_{\rm med}} P({\rm SFR}) d{\rm SFR} = 0.5, 
\label{eq:SFR_med}
\end{equation}
where $M_{\star, {\rm min}}$ and ${\rm SFR}_{\rm min}$ are the minimum values of the stellar mass and the SFR in the SED-fitting procedure, and $P(\Mstar)$ and $P({\rm SFR})$ are the PDFs of each parameter derived by marginalizing over all the other fitting parameters, respectively. 
It is noted that these physical parameters are also computed for $1\sigma$ confidence intervals. 
Hereafter, we use $\zbest$ and median values of $\Mstar$ and SFR for the physical quantities of each galaxy and just quote $z$, $\Mstar$, and SFR to represent them. 
Our galaxy samples are resampled according to their stellar mass and SFR, and details of each redshift bin are summarized in Table~\ref{tab:zbin} for stellar-mass limited subsamples and Table~\ref{tab:zbin_sfr} for SFR-limited subsamples. 

\begin{figure*}[tbp]
\plotone{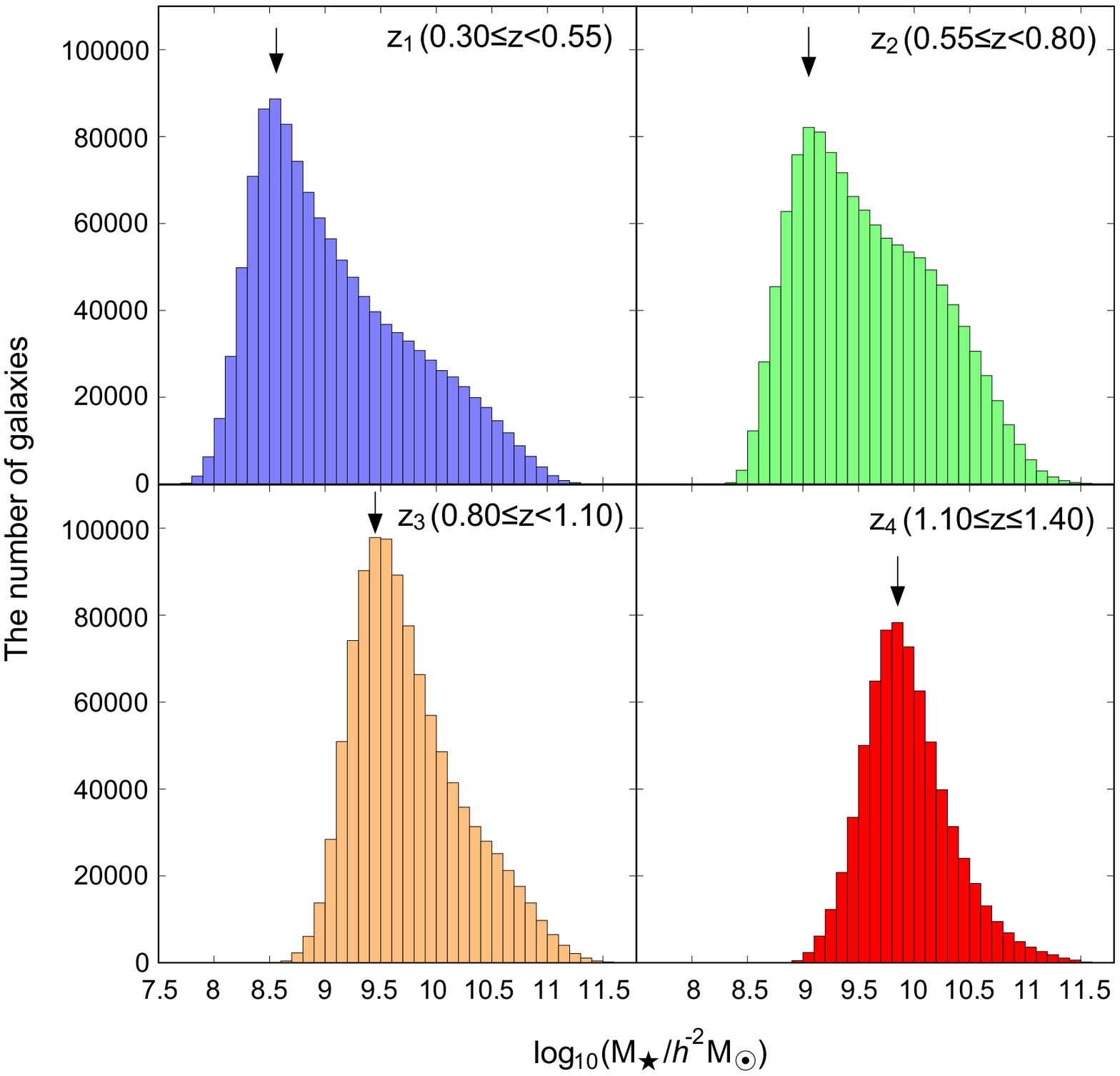}
\caption{Stellar-mass distributions of each redshift bin. The black arrows above the histogram indicate the stellar-mass limit of each redshift bin. Stellar masses are computed via the SED-fitting technique and expressed in unit of $h^{-2}\Msun$ in a logarithmic scale.}
\label{fig:Mstar_dist}
\end{figure*}

\begin{figure*}[tpb]
\plotone{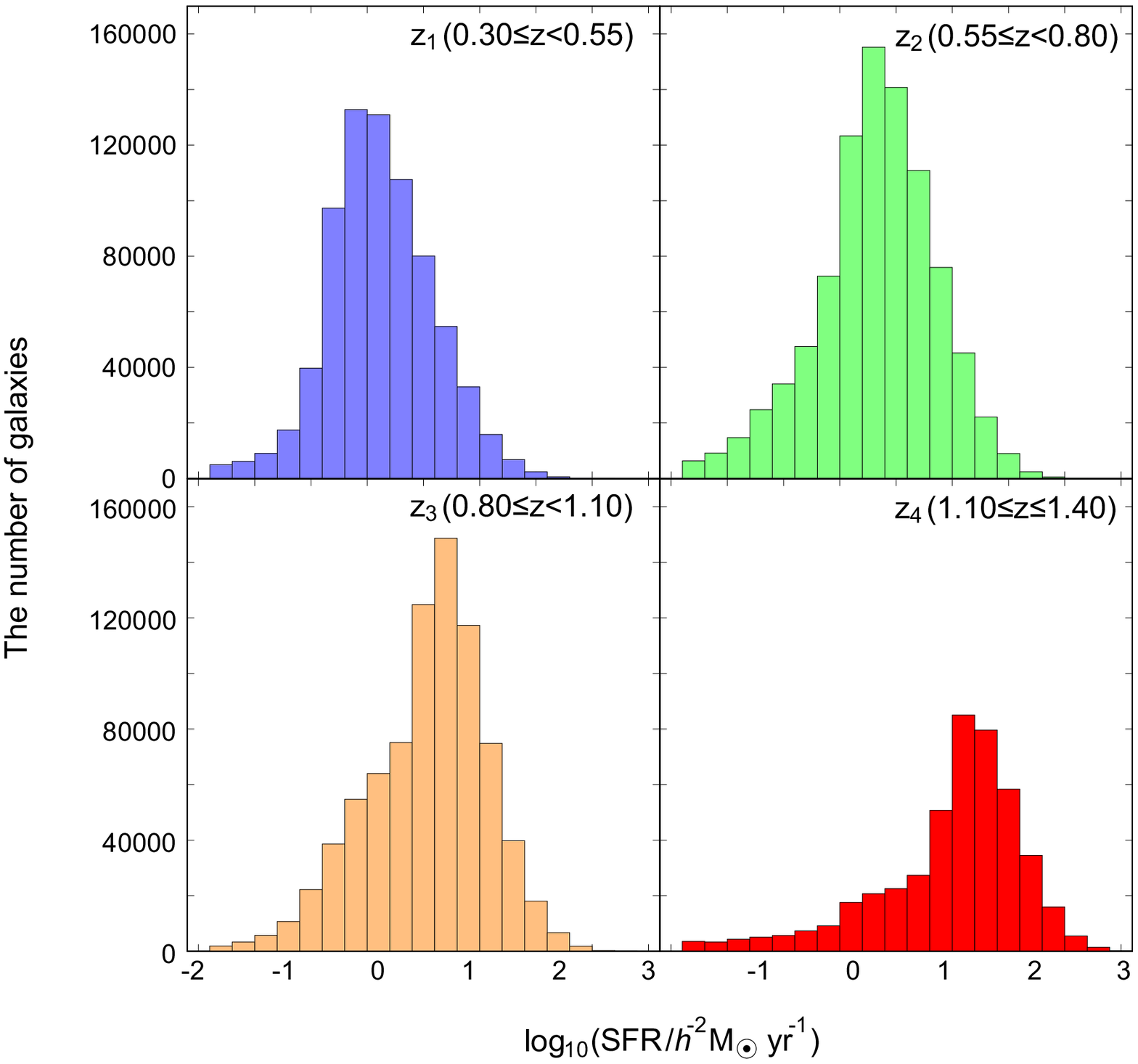}
\caption{Star-formation rate (SFR) distributions of each redshift bin. SFRs are expressed in unit of $h^{-2}\Msun$ yr$^{-1}$ in a logarithmic scale. All the galaxies satisfy the stellar-mass limit of their redshift bins. }
\label{fig:sfr_dist}
\end{figure*}

In each redshift bin, stellar-mass limits are determined as the $70\%$ stellar-mass completeness compared to the stellar-mass functions in the COSMOS field which are calculated using a publicly available COSMOS/UltraVISTA $K_{s}$-selected photometric-redshift catalogue \citep{muzzin13}. 
It should be noted that the SFR-resampled bins also satisfy the stellar-mass limit of each redshift bin. 
The stellar mass and the SFR distributions of each redshift bin are presented in Figure~\ref{fig:Mstar_dist} and \ref{fig:sfr_dist}. 

\section{Methodologies: Clustering and HOD Analysis} \label{sec:clustering}
\subsection{Angular Correlation Function} \label{subsec:acf}
We measure angular two-point auto-correlation functions (ACFs) to quantitatively estimate the strength of galaxy clustering \citep{totsuji69,peebles80}. 
The ACFs, $\omega(\theta)$, are calculated using an estimator proposed by \citet{landy93}: 
\begin{equation}
\omega\left(\theta\right) = \frac{{\rm DD} - 2 {\rm DR} + {\rm RR}}{{\rm RR}}, 
\label{eq:acf}
\end{equation}
where ${\rm DD}$, ${\rm DR}$, and ${\rm RR}$ are the number of normalized pairs of galaxy--galaxy, galaxy--random, and random--random with separation angle of $\theta \pm \delta\theta$, respectively. 
In this study, the angular scale is in unit of degree and the angular range is $-3.4 \leq \log_{10} (\theta/{\rm degree}) \leq 0.2$ with $0.2$ dex separations. 
Random samples are generated over the same region of the galaxy samples. 
To minimize the Poisson noise, we generate at least $100$ times more homogeneous random samples compared to galaxy samples over real galaxy distributions on each patch. 

The ACFs can be approximated by a power-law form as: 
\begin{equation}
\omega\left(\theta\right) = A_{\omega} \theta^{1 - \gamma}, 
\label{eq:acf_powerlaw}
\end{equation}
where $A_{\omega}$ and $\gamma$ are the amplitude of the ACF and the power-law slope, respectively. 
It is known that the ACFs of \citeauthor{landy93} are underestimated due to the limitation of the survey field, especially for large-angular scales, called ``integral constraint (IC)''. 
The integral constraint, which corresponds to the underestimation of observed ACFs from the approximated power law at large-angular scales, can be calculated as: 
\begin{equation}
{\rm IC} = \frac{\sum_{i}\theta^{1-\gamma}_{i}{\rm RR\left(\theta_{i}\right)}}{\sum_{i} {\rm RR\left(\theta_{i}\right)}}, 
\label{eq:ic}
\end{equation}
where $\theta_{i}$ is the $i$th angular scale, and ${\rm RR\left(\theta_{i}\right)}$ is the number of random--random pairs of $\theta_{i}$ bin \citep{roche99}. 
We first derive the power-law slope of each ACF by fitting the power law (equation~\ref{eq:acf_powerlaw}) for varying $A_{\omega}$ and $\gamma$, and then evaluate the integral constraints. 
However, the survey area of the HSC SSP is sufficiently wide to accurately compute ACFs even at large-angular scales $(\theta \sim 1^{\circ})$, so that the values of the integral constraints are quite small (IC $\sim 0.5$).

The errors of the ACF are estimated by computing the covariance matrix \citep{norberg09}, which is calculated using a ``delete-one'' jackknife method \citep{shao89}. 
We divide our survey field into $157$ subfields and evaluate the ACF by excluding one subfield. 
This procedure is repeated $157$ times by changing the excluded subfield. 
The covariance matrix is calculated as follows: 
\begin{equation}
C_{ij} = \frac{N-1}{N} \sum_{k=1}^{N} \left( \omega_{k}\left(\theta_{i}\right) -  \bar{\omega} \left(\theta_{i}\right) \right) \times \left( \omega_{k}\left(\theta_{j}\right) -  \bar{\omega} \left(\theta_{j}\right) \right), 
\label{eq:cm}
\end{equation}
where $C_{ij}$ is an $(i, j)$ element of the covariance matrix, $ \omega_{k}\left(\theta_{i}\right)$ is the ACF of the $i$th angular bin of the $k$th jackknife resampling, and $\bar{\omega} \left(\theta_{i}\right)$ is the averaged ACF of the $i$th angular bin, respectively. 

\subsection{Halo Occupation Distribution Model} \label{subsec:hod}
We adopt a halo occupation distribution (HOD) formalism to interpret the observed galaxy clustering. 
In the standard HOD model, physical characteristics, including the galaxy occupation, depend only on the dark halo mass; however, recent studies have reported that the clustering strengths of dark haloes also depend on their age, concentration, and spin \citep[e.g.,][]{gao05,dalal08,lacerna11}, and the large-scale environment and the formation time of host dark haloes affect the characteristics of galaxies \citep[e.g.,][]{zehavi18,artale18}, known as a {\it halo assembly bias}. 
Some advanced HOD models have been proposed to consider the effect of the halo assembly bias effect \citep[e.g.,][]{hearin16}. 
However, we employ the standard HOD model, which does not consider the halo bias parameters to achieve a fairly comparison of our results with previous clustering studies. 

The HOD model assumes that the number of galaxies within dark haloes is a function of the dark halo mass. 
The expected total number of galaxies within dark haloes with mass of $\Mh$, $N\left(\Mh\right)$, can be written as,
\begin{equation}
N\left(\Mh \right) = N_{\rm c}\left(\Mh\right) \times \left[ 1 + N_{\rm s}\left(\Mh\right) \right],
\label{eq:hof}
\end{equation}
where $N_{\rm c}\left(\Mh\right)$ and $N_{\rm s}\left(\Mh\right)$ represent the expectation number of central and satellite galaxies within dark haloes with mass of $\Mh$, respectively. 
We adopt a standard halo occupation model proposed by \citet{zheng05}:
\begin{equation}
N_{\rm c}\left(\Mh\right) = \frac{1}{2} \left[ 1 + {\rm erf}\left(\frac{\log_{10}{\Mh} - \log_{10}{\Mmin}}{\sigmaM}\right) \right],
\label{eq:Nc}
\end{equation}
and
\begin{equation}
N_{\rm s}\left(\Mh\right) = \left(\frac{\Mh - M_{0}}{M_{1}}\right)^{\alpha} ({\rm for} \: \Mh>M_{0}). 
\label{eq:Ns}
\end{equation}
In this parameterization of galaxy occupation, there are five free HOD parameters: $\Mmin$, $M_{1}$, $M_{0}$, $\sigmaM$, and $\alpha$, and a physical interpretation can be performed for each parameter. 
The central galaxy occupation is controlled by two parameters: $\Mmin$ is the dark halo mass at which half of the dark haloes possess a central galaxy, and $\sigmaM$ is the deviation from the step-function-like occupation of central galaxies. 
 Three other parameters control the occupation of satellite galaxies: $M_{1}$ is a typical dark halo mass that consists of one satellite galaxy, $M_{0}$ is the threshold dark halo mass for the formation of satellite galaxies, and $\alpha$ is the formation efficiency of satellite galaxies, respectively. 

For the physical characteristics of dark haloes, we employ the following analytical formulae: the halo density profile is assumed to be an NFW profile \citep{navarro97}, the halo mass function is a Sheth~\&~Tormen mass function \citep{sheth99,sheth01}, the large-scale halo bias is a model constructed by \citet{tinker10} with a scale-dependent halo bias relation of \citet{tinker05}, and the halo mass--concentration relation is assumed the model proposed by \citet{takada02,takada03}. 
The non-linear matter power spectrum is computed by the prescription presented by \citet{smith03} with the transfer function proposed by \citet{eisenstein98}. 
Photometric redshifts are used for redshift distributions of the galaxy samples. 
It is noted that the halo exclusion effect is considered in the prediction of ACFs from the HOD formalism \citep{zheng04,tinker05}. 

By investigating the best-fitting parameters and constraining the $1\sigma$ confidence intervals, we can derive further information about galaxies and their host dark haloes. 
The number density of galaxies from the HOD model, $n_{g}^{\rm HOD}$, can be computed using the halo occupation function (equation~\ref{eq:hof}) as follows: 
\begin{equation}
n_{g}^{\rm HOD} = \int d\Mh \frac{dn}{d\Mh} N\left( \Mh \right).
\label{eq:ng_hod}
\end{equation}
Here, $\frac{dn}{d\Mh}$ is the \citeauthor{sheth99} halo mass function. 
The satellite fraction, $\fsat$, which represents the percentage of the satellite galaxies to the total galaxies can be evaluated as:
\begin{eqnarray}
\fsat & = & 1 - f_{\rm c} \nonumber \\
 & = & 1 - \frac{1}{n_{g}^{\rm HOD}} \int d\Mh \frac{dn}{d\Mh} N_{\rm c}\left( \Mh \right), 
\label{eq:fs}
\end{eqnarray}
where $f_{\rm c}$ denotes the fraction of central galaxies. 
The large-scale galaxy bias, $\bg$, that associates galaxies with the underlying dark matter distribution \citep[][for a reveiw]{desjacques16} is computed as: 
\begin{equation}
\bg = \frac{1}{n_{g}^{\rm HOD}} \int d\Mh \bh\left(\Mh\right) \frac{dn}{d\Mh} N\left( \Mh \right), 
\label{eq:bgal}
\end{equation}
where $\bh \left(\Mh\right)$ is the large-scale halo bias \citep{tinker10}. 
In addition to the above large-scale galaxy bias derived by the best-fitting HOD parameters, one can calculate the galaxy bias using another definition as follows: 
\begin{equation}
\bg = \sqrt{\frac{\xi_{gg}}{\xi_{DM}}}, 
\label{eq:bgal2}
\end{equation}
where $\xi_{gg}$ and $\xi_{DM}$ are spatial auto-correlation functions of galaxies and dark matter, respectively. 
\citet{zehavi11} reported that the large-scale galaxy bias calculated using both methods are almost comparable, and we employ the large-scale galaxy bias based on the HOD modeling (equation~\ref{eq:bgal}) in the following analyses. 

\section{Results: Clustering and HOD-model analysis} \label{sec:result}
\subsection{Measurement of Angular Auto-correlation Functions} \label{subsec:acf}
We measure the ACFs of both stellar-mass limited samples (hereafter, SM samples) and SFR-limited samples (hereafter, SFR samples) and they are shown in Figure~\ref{fig:HOD_Mstar} and \ref{fig:HOD_sfr}, respectively. 
The observed ACFs are well approximated by a power law at a large-angular scale, since the HSC SSP survey covers a sufficiently wide area to precisely calculate degree-scale clustering. 
At a small-angular scale, the apparent excesses from the power law known as a $1$-halo term is evident. 
The $1$-halo term is prominent for massive stellar-mass bins regardless of its redshift. 

The measured ACFs are fitted using a power law (equation~\ref{eq:acf_powerlaw}) after IC correction. 
The high S/N ACFs based on a wide survey field of the HSC SSP allow the amplitude of ACFs and the power-law slope to be simultaneously constrained. 
We present the fitting results for the amplitudes of ACFs at $1^{\circ}$ and the power-law slopes of ACFs at a large-angular scale for each subsample in Table~\ref{tab:Mstar_clustering} and \ref{tab:sfr_clustering}. 
The amplitudes of galaxy clustering increase monotonically with the stellar-mass limit. 
However, the power-law slope of ACFs, $\gamma$, is almost constant as $\gamma \sim 1.7-1.8$, although less-massive subsamples tend to have shallower power-law slopes compared to massive subsamples. 
We investigate and discuss the dependence of clustering strengths on galaxy stellar masses and SFRs by measuring correlation lengths in Section~\ref{subsec:r0}. 

\begin{table*}[tbp]
\begin{center}
\caption{Clustering Properties of Cumulative Stellar-mass Limited Subsamples}
\begin{tabular}{lccccccc} \hline \hline
 &  & $z_{1}$ & & & & $z_{2}$ &  \\ 
& \multicolumn{3}{c}{$0.30 \leq z < 0.55$} & & \multicolumn{3}{c}{$0.55 \leq z < 0.80$} \\
\cline{2-4}\cline{6-8}
Stellar-mass limit\footnote{Threshold stellar mass of each subsample in units of $h^{-2}\Msun$ in a logarithmic scale. } & \multicolumn{1}{c}{$A_{\omega}(\times 10^{-3})$\footnote{Amplitude of the ACF at $1^{\circ}$. }} & \multicolumn{1}{c}{$\gamma$\footnote{Power-law slope of the ACF at large-angular scale. }} & \multicolumn{1}{c}{$r_{0}$\footnote{Correlation length in units of $h^{-1}$Mpc. }} & & \multicolumn{1}{c}{$A_{\omega}(\times 10^{-3})$} & \multicolumn{1}{c}{$\gamma$} & \multicolumn{1}{c}{$r_{0}$}  \\ \hline
$8.6$ & $7.21 \pm 1.31$ & $1.79 \pm 0.09$ & $3.94 \pm 0.31$ & & -- & -- & -- \\
$8.8$ & $7.71 \pm 1.34$ & $1.80 \pm 0.09$ & $4.11 \pm 0.21$ & & -- & -- & -- \\
$9.0$ & $8.22 \pm 1.42$ & $1.81 \pm 0.09$ & $4.29 \pm 0.29$ & & $7.04 \pm 1.38$ & $1.78 \pm 0.09$ & $4.17 \pm 0.27$ \\
$9.2$ & $8.91 \pm 1.59$ & $1.81 \pm 0.09$ & $4.62 \pm 0.43$ & & $6.87 \pm 1.36$ & $1.81 \pm 0.09$ & $4.20 \pm 0.25$ \\
$9.4$ & $9.49 \pm 1.75$ & $1.81 \pm 0.10$ & $4.67 \pm 0.35$ & & $7.13 \pm 1.33$ & $1.82 \pm 0.08$ & $4.28 \pm 0.22$ \\
$9.6$ & $10.39 \pm 1.99$ & $1.80 \pm 0.10$ & $4.85 \pm 0.22$ & & $7.22 \pm 1.31$ & $1.83 \pm 0.08$ & $4.67 \pm 0.10$ \\
$9.8$ & $11.20 \pm 2.14$ & $1.80 \pm 0.10$ & $4.96 \pm 0.22$ & & $7.75 \pm 1.44$ & $1.82 \pm 0.08$ & $5.04 \pm 0.19$ \\
$10.0$ & $11.96 \pm 2.19$ & $1.81 \pm 0.10$ & $5.19 \pm 0.40$ & & $8.45 \pm 1.55$ & $1.82 \pm 0.08$ & $5.16 \pm 0.19$ \\
$10.2$ & $12.86 \pm 2.56$ & $1.80 \pm 0.10$ & $5.58 \pm 0.47$ & & $9.97 \pm 1.88$ & $1.80 \pm 0.09$ & $5.67 \pm 0.29$ \\
$10.4$ & $14.66 \pm 3.80$ & $1.79 \pm 0.10$ & $6.29 \pm 0.53$ & & $12.40 \pm 2.57$ & $1.77 \pm 0.10$ & $6.07 \pm 0.33$ \\
$10.6$ & $17.88 \pm 6.57$ & $1.78 \pm 0.10$ & $6.68 \pm 0.57$ & & $15.18 \pm 4.13$ & $1.77 \pm 0.10$ & $6.32 \pm 0.48$ \\
$10.8$ & $27.53 \pm 16.31$ & $1.78 \pm 0.10$ & $7.56 \pm 0.80$ & & $17.94 \pm 7.54$ & $1.78 \pm 0.10$ & $6.84 \pm 0.65$ \\
$11.0$ & $33.19 \pm 21.11$ & $1.79 \pm 0.11$ & $9.21 \pm 1.04$ & & $25.59 \pm 14.74$ & $1.78 \pm 0.11$ & $8.07 \pm 0.97$ \\
$11.2$ & -- & -- & -- & & $38.33 \pm 25.82$ & $1.78 \pm 0.10$ & $9.60 \pm 1.04$ \\ \hline
\end{tabular}

\vspace{5pt}

\begin{tabular}{lccccccc}
 &  & $z_{3}$ & & & & $z_{4}$ &  \\ 
& \multicolumn{3}{c}{$0.80 \leq z < 1.10$} & & \multicolumn{3}{c}{$1.10 \leq z \leq 1.40$} \\
\cline{2-4}\cline{6-8}
Stellar-mass limit & \multicolumn{1}{c}{$A_{\omega}(\times 10^{-3})$} & \multicolumn{1}{c}{$\gamma$} & \multicolumn{1}{c}{$r_{0}$} & & \multicolumn{1}{c}{$A_{\omega}(\times 10^{-3})$} & \multicolumn{1}{c}{$\gamma$} & \multicolumn{1}{c}{$r_{0}$}  \\ \hline
$9.4$ & $4.67 \pm 0.99$ & $1.83 \pm 0.08$ & $4.33 \pm 0.26$ & & -- & -- & -- \\
$9.6$ & $5.29 \pm 1.10$ & $1.83 \pm 0.08$ & $4.45 \pm 0.26$ & & -- & -- & -- \\
$9.8$ & $6.03 \pm 1.20$ & $1.83 \pm 0.08$ & $4.59 \pm 0.34$ & & $6.64 \pm 1.36$ & $1.75 \pm 0.09$ & $3.87 \pm 0.17$ \\
$10.0$ & $7.48 \pm 1.63$ & $1.81 \pm 0.08$ & $5.13 \pm 0.41$ & & $6.97 \pm 1.45$ & $1.76 \pm 0.09$ & $4.11 \pm 0.32$ \\
$10.2$ & $9.03 \pm 2.26$ & $1.80 \pm 0.09$ & $5.80 \pm 0.47$ & & $7.42 \pm 1.83$ & $1.77 \pm 0.10$ & $4.60 \pm 0.40$ \\
$10.4$ & $11.09 \pm 2.82$ & $1.78 \pm 0.10$ & $6.29 \pm 0.33$ & & $7.75 \pm 2.24$ & $1.78 \pm 0.10$ & $5.09 \pm 0.54$ \\
$10.6$ & $13.24 \pm 4.64$ & $1.78 \pm 0.10$ & $6.68 \pm 0.72$ & & $8.56 \pm 4.31$ & $1.78 \pm 0.10$ & $6.86 \pm 0.64$ \\
$10.8$ & $15.32 \pm 6.88$ & $1.77 \pm 0.10$ & $6.95 \pm 0.84$ & & $10.60 \pm 6.77$ & $1.78 \pm 0.10$ & $8.25 \pm 0.49$ \\
$11.0$ & $25.24 \pm 16.17$ & $1.77 \pm 0.10$ & $8.56 \pm 1.03$ & & $22.23 \pm 15.61$ & $1.77 \pm 0.10$ & $8.70 \pm 1.07$ \\ 
$11.2$ & $33.14 \pm 21.60$ & $1.78 \pm 0.10$ & $10.52 \pm 1.26$ & & $41.88 \pm 26.22$ & $1.78 \pm 0.10$ & $9.97 \pm 1.06$ \\ \hline
\end{tabular}
\label{tab:Mstar_clustering}
\end{center}
\end{table*}

\begin{table*}[tbp]
\begin{center}
\caption{Clustering Properties of Cumulative Star-formation Rate Limited Subsamples}
\begin{tabular}{lccccccc} \hline \hline
 &  & $z_{1}$ & & & & $z_{2}$ &  \\ 
& \multicolumn{3}{c}{$0.30 \leq z < 0.55$} & & \multicolumn{3}{c}{$0.55 \leq z < 0.80$} \\
\cline{2-4}\cline{6-8}
SFR limit\footnote{Threshold SFR of each subsample in units of $h^{-2}\Msun$yr$^{-1}$ in a logarithmic scale} & \multicolumn{1}{c}{$A_{\omega}(\times 10^{-3})$\footnote{Amplitude of the ACF at $1^{\circ}$}} & \multicolumn{1}{c}{$\gamma$\footnote{Power-law slope of the ACF at large-angular scale}} & \multicolumn{1}{c}{$r_{0}$\footnote{Correlation length in units of $h^{-1}$Mpc}} & & \multicolumn{1}{c}{$A_{\omega}(\times 10^{-3})$} & \multicolumn{1}{c}{$\gamma$} & \multicolumn{1}{c}{$r_{0}$}  \\ \hline
$-1.50$ & $6.44 \pm 1.36$ & $1.73 \pm 0.08$ & $3.22 \pm 0.06$ & & $7.11 \pm 1.31$ & $1.73 \pm 0.07$ & $3.73 \pm 0.17$ \\
$-1.00$ & $6.44 \pm 1.32$ & $1.72 \pm 0.08$ & $3.22 \pm 0.12$ & & $6.72 \pm 1.28$ & $1.73 \pm 0.07$ & $3.68 \pm 0.25$ \\
$-0.50$ & $6.80 \pm 1.52$ & $1.72 \pm 0.08$ & $3.49 \pm 0.13$ & & $6.12 \pm 1.26$ & $1.74 \pm 0.08$ & $3.69 \pm 0.25$ \\
$0.00$ & $8.28 \pm 2.57$ & $1.74 \pm 0.09$ & $3.81 \pm 0.16$ & & $5.22 \pm 1.30$ & $1.78 \pm 0.09$ & $3.91 \pm 0.26$ \\
$0.50$ & $19.61 \pm 11.29$ & $1.77 \pm 0.10$ & $5.33 \pm 0.49$ & & $5.77 \pm 2.54$ & $1.77 \pm 0.10$ & $4.85 \pm 0.43$ \\ \hline
\end{tabular}

\vspace{5pt}

\begin{tabular}{lccccccc}
 &  & $z_{3}$ & & & & $z_{4}$ &  \\ 
& \multicolumn{3}{c}{$0.80 \leq z < 11.0$} & & \multicolumn{3}{c}{$11.0 \leq z \leq 1.40$} \\
\cline{2-4}\cline{6-8}
SFR limit & \multicolumn{1}{c}{$A_{\omega}(\times 10^{-3})$} & \multicolumn{1}{c}{$\gamma$} & \multicolumn{1}{c}{$r_{0}$} & & \multicolumn{1}{c}{$A_{\omega}(\times 10^{-3})$} & \multicolumn{1}{c}{$\gamma$} & \multicolumn{1}{c}{$r_{0}$}  \\ \hline
$-1.00$ & $4.77 \pm 1.01$ & $1.83 \pm 0.09$ & $3.83 \pm 0.10$ & & -- & -- & -- \\
$-0.50$ & $4.49 \pm 0.97$ & $1.83 \pm 0.09$ & $3.77 \pm 0.17$ & & $6.60 \pm 1.35$ & $1.73 \pm 0.08$ & $3.81 \pm 0.17$ \\
$0.00$ & $4.43 \pm 1.00$ & $1.81 \pm 0.10$ & $3.89 \pm 0.36$ & & $6.11 \pm 1.44$ & $1.75 \pm 0.09$ & $3.95 \pm 0.45$ \\
$0.50$ & $4.68 \pm 1.88$ & $1.79 \pm 0.10$ & $4.77 \pm 0.41$ & & $5.83 \pm 1.41$ & $1.76 \pm 0.09$ & $4.18 \pm 0.62$ \\
$1.00$ & $15.28 \pm 10.79$ & $1.78 \pm 0.11$ & $7.30 \pm 0.82$ & & $7.26 \pm 3.28$ & $1.77 \pm 0.10$ & $5.11 \pm 1.13$ \\ \hline
\end{tabular}
\label{tab:sfr_clustering}
\end{center}
\end{table*}

\subsection{HOD-model Fitting} \label{subsec:hod_fitting}
We implement the HOD-model fitting to interpret the observed HSC galaxy clustering. 
In the standard HOD model presented in Section~\ref{subsec:hod}, there are five free HOD parameters. 
By comparing the observed ACFs with those predicted by the HOD model, we investigate the best-fitting HOD parameters to represent the observed galaxy clustering signals that originate from the environmental and physical properties of the residing dark haloes. 

\subsubsection{The HOD fitting on stellar-mass samples} \label{subsubsec:hod_Mstar}
According to the stellar mass--halo mass relation \citep{leauthaud12,matthee17}, massive dark haloes tend to possess massive central galaxies. 
Therefore, in the cumulative SM samples, less massive dark halos can stochastically possess less massive central galaxies, whereas massive dark haloes should have one massive central galaxy and several satellite galaxies. 
This situation can be treated by the halo occupation function of the HOD formalism. 

The best-fitting parameters are evaluated through the least $\chi^{2}$ fitting procedure as follows: 
\begin{eqnarray}
\chi^{2} & = & \sum_{i, j} \left[ \omega^{\rm obs}\left(\theta_{i}\right) - \omega^{\rm HOD}\left(\theta_{i}\right) \right] \left( C_{ij}^{-1} \right) \left[ \omega^{\rm obs}\left(\theta_{j}\right) - \omega^{\rm HOD}\left(\theta_{j}\right) \right] \nonumber \\
& + & \frac{\left( n_{g}^{\rm obs} - n_{g}^{\rm HOD} \right)^{2}}{{\sigma_{n_{g}}}^{2}}, 
\label{eq:chi_hod}
\end{eqnarray}
where $\omega^{\rm obs}\left(\theta_{i}\right)$ and $\omega^{\rm HOD}\left(\theta_{i}\right)$ are the ACFs of $i$th angular bin calculated by observation and the HOD model, $C_{ij}^{-1}$ is an $(i, j)$ element of an inverse covariance matrix (equation~\ref{eq:cm}), $n_{g}^{\rm obs}$ and $n_{g}^{\rm HOD}$ are the number density of galaxies computed by observation and the HOD model (equation~\ref{eq:ng_hod}), and $\sigma_{n_{g}}$ is an uncertainty of the observed number density of galaxies, respectively. 
The correlation factor presented by \citet{hartlap07} is introduced to obtain an unbiased inverse covariance matrix. 
We consider the uncertainty of the number of galaxy samples due to errors associated with photometric redshifts as well as the selection incompleteness. 
The contamination fractions of each redshift bin according to the photometric redshift were measured using galaxies that exhibited spectroscopic redshifts \citep{tanaka18}, and we include the effect of contaminations in the uncertainty of the number of galaxies (for example, $12\%$ uncertainties for $z_{1}$ bin and $25\%$ for $z_{4}$ bin). 
In addition, selection incompleteness is considered in the uncertainty of galaxy number density; therefore, the HOD parameters of less-massive subsamples cannot be strongly constrained due to the relatively large uncertainties of the galaxy abundance. 

\begin{figure*}[tbp]
\plotone{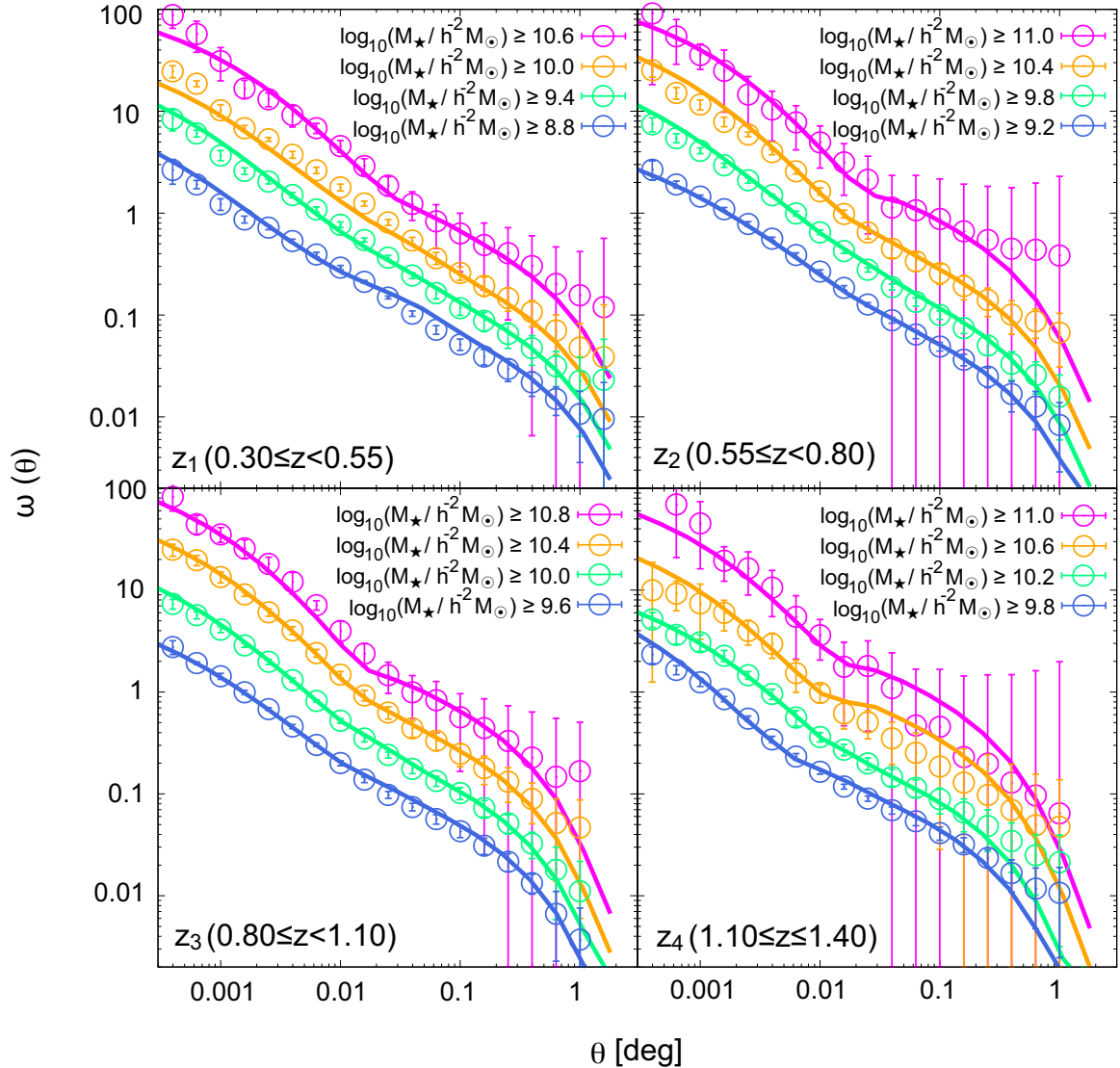}
\caption{The observed ACFs of the SM samples (circles) and their best-fitting ACFs derived using the HOD model (solid lines) at $0.30 \leq z < 0.55$ (top left panel), $0.55 \leq z < 0.80$ (top right panel), $0.80 \leq z < 1.10$ (bottom left panel), and $1.10 \leq z \leq 1.40$ (bottom right panel) redshift bins, respectively. In the HOD-model fitting procedure, the correlations between the angular bins are considered using the covariance matrices based on the jackknife resampling method. For clarity, amplitudes of ACFs are normalized arbitrarily. }
\label{fig:HOD_Mstar}
\end{figure*}

Part of the results of the HOD parameter fitting on SM samples are presented in Figure~\ref{fig:HOD_Mstar}. 
One can find that the HOD model obviously succeeds in reconstructing the observed galaxy clustering for all stellar-mass ranges and wide dynamical scales at $0.3 \leq z \leq 1.4$. 

As shown in Section~\ref{subsec:hod}, the physical quantities of the host dark haloes can be evaluated using the best-fitting HOD parameters. 
We discuss the details and the redshift evolution of the host dark haloes of the HSC galaxy samples in the following sections. 

\subsubsection{The HOD fitting on SFR samples} \label{subsubsec:hod_sfr}
\begin{figure*}[tbp]
\plotone{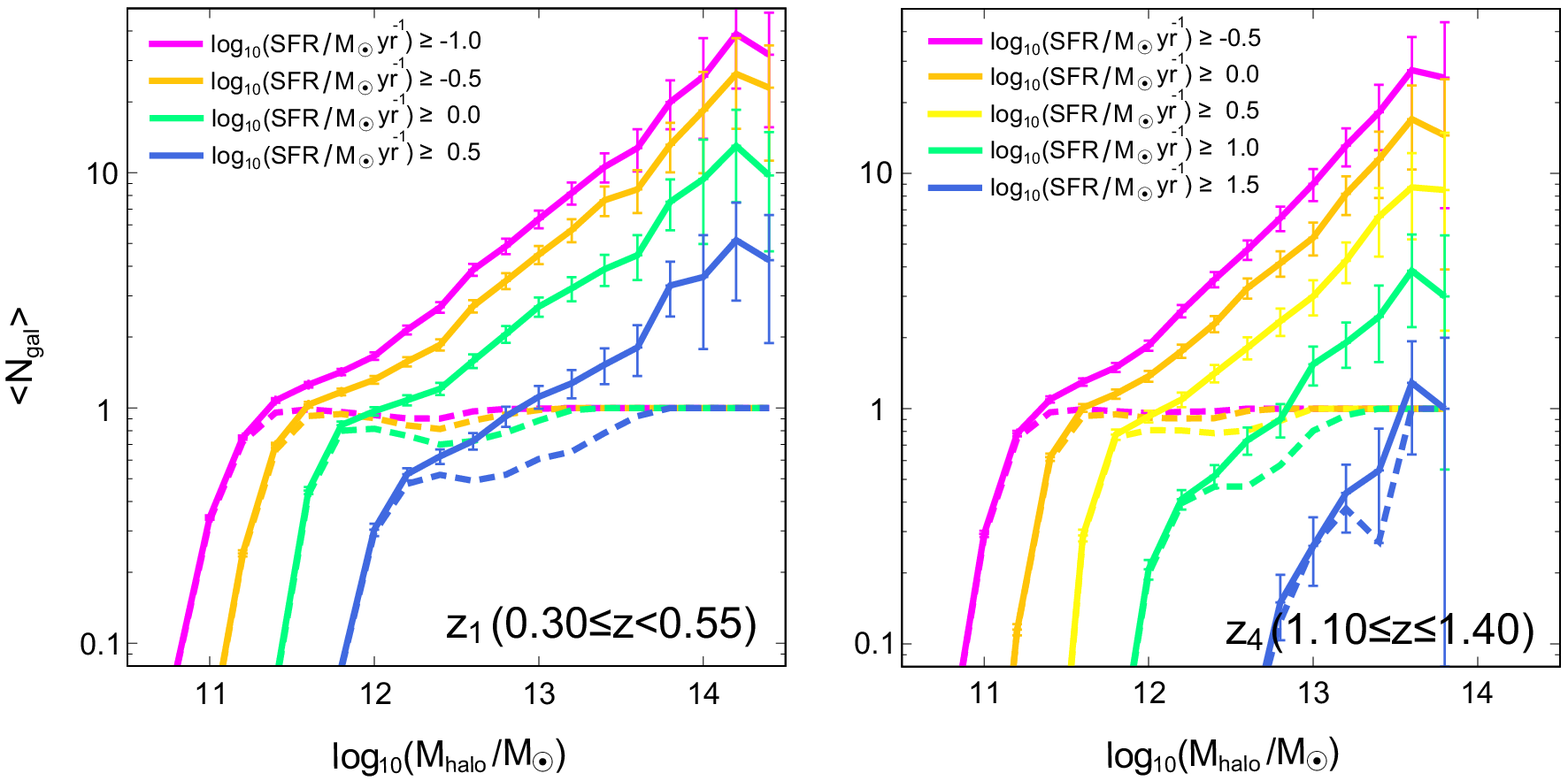}
\caption{The expected number of galaxies per dark haloes calculated using the data from the EAGLE simulation. Left (right) panel shows the result for the galaxies at $0.30\leq z < 0.55$ ($1.10 \leq z \leq 1.40$) and the dashed (solid) lines represent the expected number of central (total) galaxies. All the galaxies are satisfied the magnitude and stellar-mass limits that are imposed on the HSC galaxy samples. The errors of total galaxy number are evaluated using the Poisson error. }
\label{fig:sfr_hof}
\end{figure*}

In addition to the SM samples, we also implement the HOD-model analysis on the SFR samples. 
However, the occupational relation of galaxies selected according to their SFR has not been established, and the halo occupation function of equation~\ref{eq:hof} is not ensured the galaxy occupation of the SFR samples. 
To verify the halo occupation pattern of the SFR samples, we examined the expected number of galaxies per haloes as a function of the dark halo mass using the publicly available data of the EAGLE simulations \citep{schaye15,mcalpine16} and a part of results are shown in Figure~\ref{fig:sfr_hof}. 
We determined the cumulative SFR samples including low-SFR galaxies can be approximated as the same occupation function of the SM samples, although occupation of central galaxies confined to high-SFR samples has an offset compared to the error function. 
To avoid the discrepancy between the high-SFR galaxy occupation and the standard halo occupation function, we performed the HOD-model fitting only on galaxies with moderately SFR limits, assuming the halo occupation function of equation~\ref{eq:hof}. 

In the HOD analysis of the SFR samples, one should also consider the possibility of a lack of passive centrals. 
When some passive galaxies are excluded by applying a threshold to the SFR, a portion of the dark haloes do not possess their central galaxy and the expected number of galaxies within the dark haloes will never reach one. 
As discussed in \citet{bethermin14}, stochastic occupation of the central galaxy does not have an impact on the shape of the ACFs predicted by the HOD model. 
However, the calculation of the number density of galaxies using the HOD model is associated with a certain problem, given that the halo occupation function plays a role in controlling the weight of the number of galaxies per halo mass. 
To deal with the stochastic occupation of SFR samples, we exclude the constraint on the galaxy abundance in the $\chi^{2}$ fitting procedure, in a similar manner to \citet{bethermin14}. 

\begin{figure*}[tbp]
\plotone{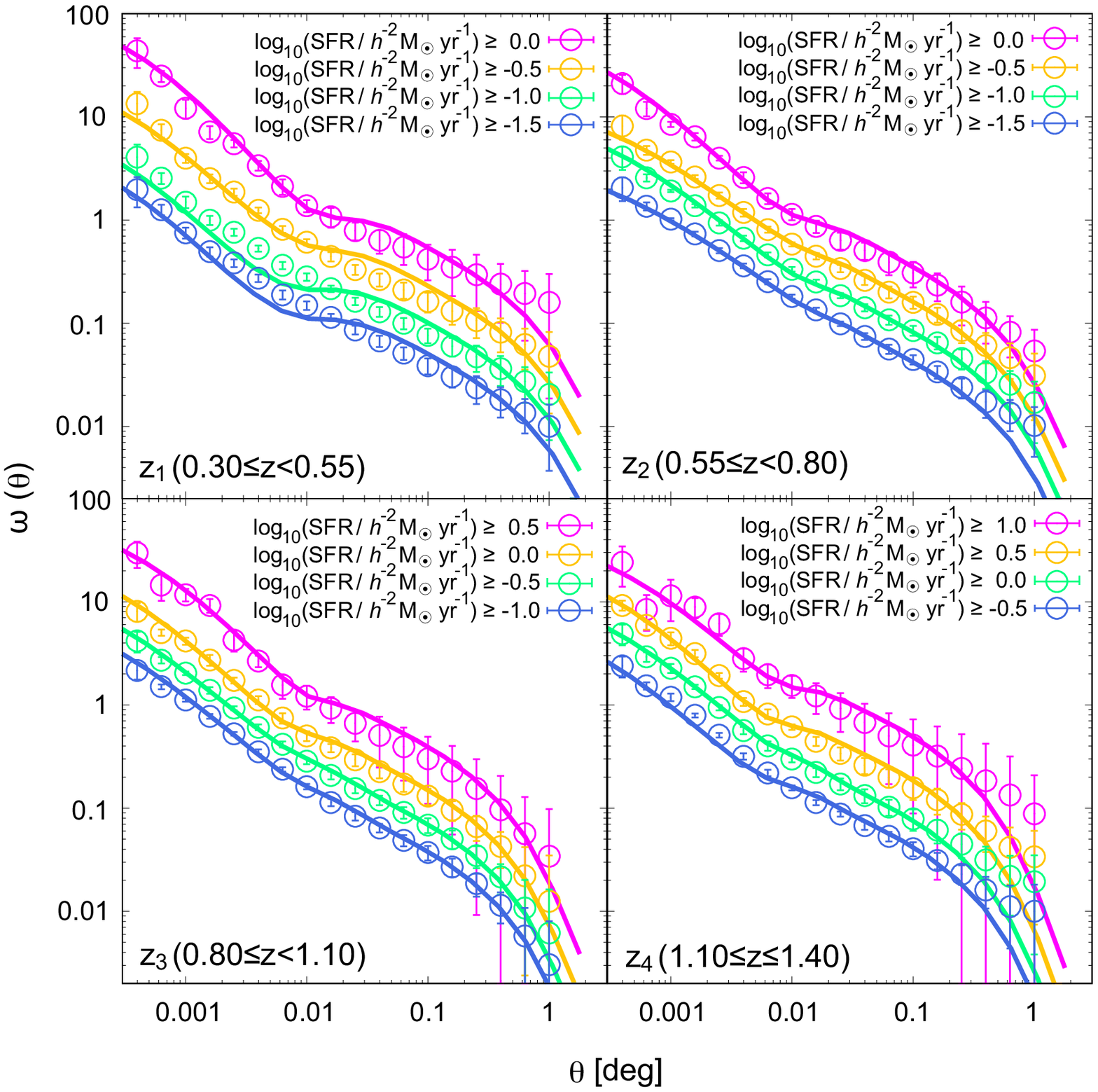}
\caption{SiImilar to Figure~\ref{fig:HOD_Mstar}, but for the SFR samples. }
\label{fig:HOD_sfr}
\end{figure*}

We show the results of the HOD-model fitting on the SFR samples in Figure~\ref{fig:HOD_sfr} and the best-fitting HOD parameters are given in Table~\ref{tab:SFR_hod}. 
As in the case with the SM samples, the HOD model can well represent observed galaxy clustering with a wide dynamical range, SFR thresholds, and redshift. 
Due to the lack of constraint on the galaxy number density, the best-fitting HOD parameters are less constrained compared to the results of SM samples. 
To precisely investigate the relationship between dark halo properties and the SFR, it is necessary to develop the halo occupation function for SFR samples and/or a sophisticated halo-model analysis method. 

\subsection{Correlation Length} \label{subsec:r0}
We measure the correlation lengths of each ACF to quantify the dependence of clustering strength on their baryonic properties. 
Correlation length is a three-dimensional clustering strength that can be evaluated by a spatial correlation function. 
The spatial two-point correlation function, $\xi(r)$, can be described in a power-law form as:
\begin{equation}
\xi(r) = \left(\frac{r}{r_{0}}\right)^{-\gamma}, 
\label{eq:real-space_cf}
\end{equation}
where the normalization factor $r_{0}$ is the correlation length and $\gamma$ is the gradient of power-law approximation of the spatial correlation function. 

Provided that the redshift distribution of galaxies is known, one can transform the amplitude of ACF into the correlation length using the ``Limber's approximation'' \citep{limber53,limber54}. 
The correlation length can be evaluated by determining the amplitude of ACFs \citep{peebles80,efstathiou91} and many clustering studies adopt this approximation formula \citep[e.g.,][]{hildebrandt09,mccracken10,ishikawa15}. 
However, the accuracy of the correlation length becomes worse with the increase of the redshift range of the galaxy sample. 
This is because the Limber's approximation converts the projected clustering amplitude into the real-space clustering strength. 

To avoid the uncertainties that originate from the Limber's approximation, we evaluate the correlation lengths by definition; i.e., the correlation length is a scale in which the real-space two-point correlation function becomes unity. 
Using the halo-model approach \citep[e.g.,][]{seljak00,ma00}, a galaxy power spectrum, $P_{\rm g}$, can be computed using the halo occupation function (equation~\ref{eq:hof}). 
To calculate the galaxy power spectrum, we use the best-fitting HOD parameters of each stellar-mass/SFR bin estimated in Section~\ref{subsec:hod_fitting}. 
The real-space two-point correlation function from the halo model can be obtained by performing a Fourier transformation of the galaxy power spectrum as follows:
\begin{equation}
\xi_{\rm HOD}\left(r\right) = \frac{1}{2\pi^{2}} \int dk k^{2} \frac{\sin{kr}}{kr} P_{\rm g}\left(k\right). 
\label{eq:wk}
\end{equation}
We determine the correlation length as a scale which satisfies $\xi_{\rm HOD}\left(r_{0}\right) = 1$. 

\subsubsection{$\Mstar$ versus $r_{0}$ relation} \label{subsubsec:Mstar-r0}
\begin{figure}[tpb]
\epsscale{1.25}
\plotone{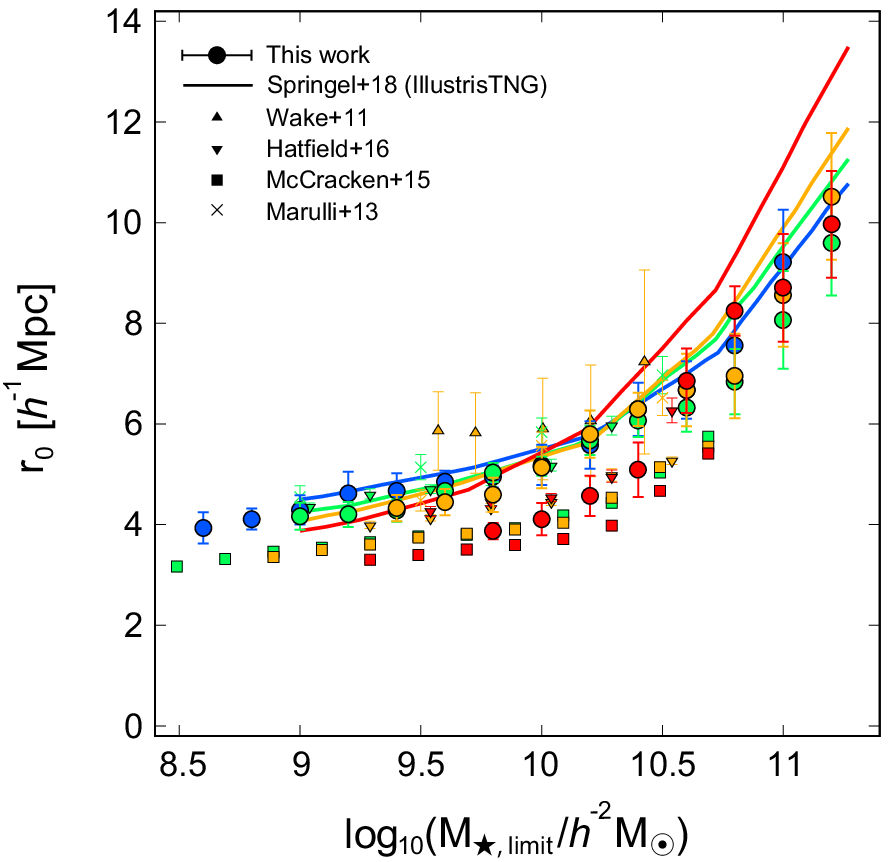}
\caption{The stellar-mass dependence and redshift evolution of the correlation length. Our results are shown by the blue ($0.30 \leq z < 0.55$), green ($0.55 \leq z < 0.80$), orange ($0.80 \leq z < 1.10$), and red ($1.10 \leq z \leq 1.40$) circles, respectively. For comparison, we plot previous observational results reported by \citet[][upward triangles]{wake11} at $0.9<z<1.3$, \citet[][downward triangles]{hatfield16} at $0.50<z<0.75$ (green), $0.75<z<1.00$ (orange), $1.00<z<1.25$ (red), \citet[][squares]{mccracken15} at $0.5<z<0.8$ (green), $0.8<z<1.1$ (orange), $1.1<z<1.5$ (red), and \citet[][crosses]{marulli13} at $0.5<z<0.7$ (green), $0.7<z<1.1$ (orange), respectively. In addition to the above observational studies, results obtained using the hydrodynamical simulations \citep[IllustrisTNG;][solid lines]{springel18} at $z=0.5$ (blue), $z=0.8$ (green), $z=1.0$ (orange), and $z=1.5$ (red) are also shown.}
\label{fig:r0}
\end{figure}

The stellar-mass dependence of the correlation length is shown in Figure~\ref{fig:r0} and the details are presented in Table~\ref{tab:Mstar_clustering}. 
Regardless of the redshift bins, the correlation lengths monotonically increase with the threshold stellar mass, which is also seen in other observational studies as well as the hydrodynamical simulations, and this trend is also observed up to $z \sim 2$ \citep[e.g.,][]{bielby14,ishikawa15}. 
This implies that massive galaxies occupy more massive dark haloes compared to less-massive galaxies. 
The correlation lengths gradually increase with the stellar-mass threshold, whereas the amplitude of the correlation lengths increase significantly when galaxy samples are confined to massive ones. 
The knee stellar-mass threshold is $\logMstarlimit \sim 10.4$ irrespective of its redshift, indicating that galaxies with larger masses than this threshold are highly biased objects. 
The steep increase of the correlation length at the massive end has already been reported in previous studies \citep[e.g.,][]{wake11,hatfield16}. 
However, we determined that it continues up to the most massive galaxies by observation. 

The redshift evolution of the correlation length is almost consistent with the results of previous studies as shown in Figure~\ref{fig:r0}. 
By fixing the stellar-mass threshold, our results indicate a small redshift evolution over $0.3 \leq z \leq 1.4$, which is also seen in the results in \citet{mccracken15} and \citet{springel18}. 
These studies measured correlation lengths without assuming the Limber's approximation. 
\citet{springel18} determined that by fixing the stellar-mass threshold, the correlation lengths are stronger for less-massive low-$z$ galaxies and this trend is reversed for massive galaxies. 
Our results support the trend of redshift evolution and show good consistency with the results obtained from the numerical hydrodynamical simulations. 
The discrepancy in the massive end between our highest-$z$ result (red circles) and the result of \citet{springel18} at $z=1.5$ (red line) is caused by the difference in the effective redshift. 
The effective redshift of our result is $z=1.23$, whereas \citet{springel18} calculated the exact result at $z=1.5$ using a snapshot at that epoch. 
\citet{mccracken15} also measured a small redshift evolution of the correlation length. 
However, their results clearly show a weaker clustering strength compared to other studies ($1-1.5\sigma$ lower than our results). 
The origin of this discrepancy is still unclear. 
The possible reason is that the COSMOS field is a relatively small survey field and it is difficult to measure the angular correlation function with sufficient dynamical range and the S/N ratio to compute an accurate real-space correlation function for the scale to evaluate the correlation length. 

\subsubsection{SFR versus $r_{0}$ relation} \label{subsubsec:sfr-r0}
\begin{figure}[tpb]
\epsscale{1.25}
\plotone{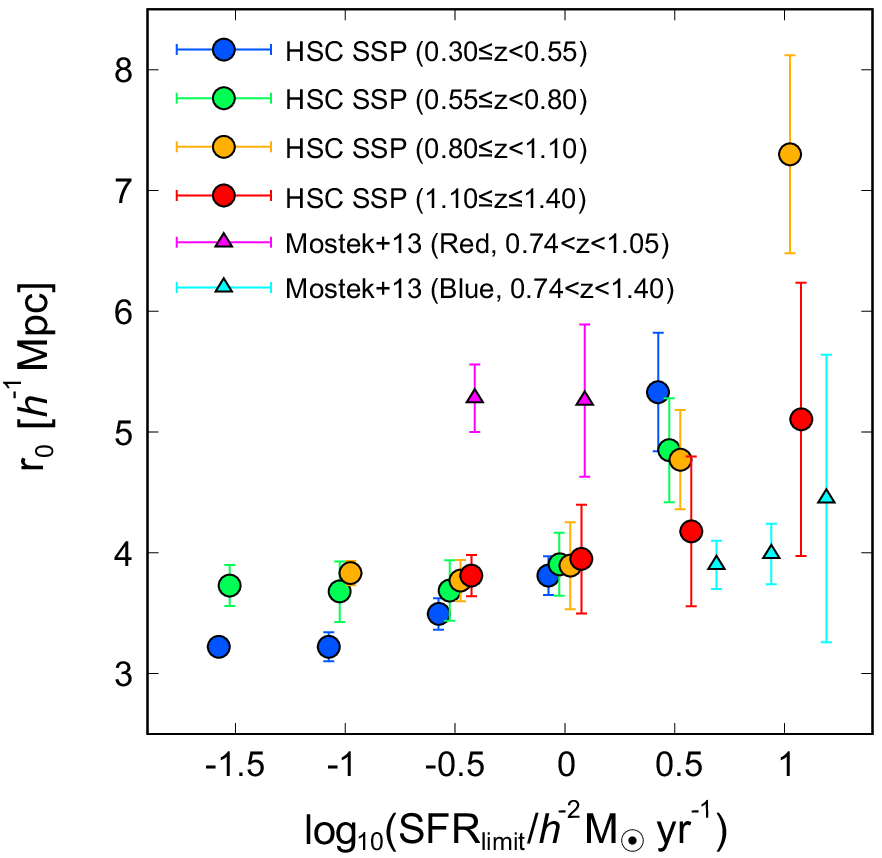}
\caption{The SFR dependence of the correlation lengths. We show our results at $0.30 \leq z < 0.55$ ($z_{1}$; blue), $0.55 \leq z < 0.80$ ($z_{2}$; green), $0.80 \leq z < 1.10$ ($z_{3}$; orange), and $1.10 \leq z \leq1.40$ ($z_{4}$; red) as a function of SFR limit. We also plot correlation lengths of red galaxies at $0.74<z<1.05$ (magenta triangles) and blue galaxies at $0.74<z<1.40$ (cyan triangles) reported by \citet{mostek13}. For clarity, our results are slightly shifted along the horizontal axis. }
\label{fig:r0_sfr}
\end{figure}

We also measure the correlation lengths of SFR samples as shown in Figure~\ref{fig:r0_sfr}. 
The correlation lengths of red and blue galaxies presented by \citet{mostek13} that are separated on the $(U-B)$ color versus $B$-band magnitude diagram are plotted for the comparison, and our correlation lengths are consistent with the results of the blue galaxies. 
Increasing the correlation length with the SFR is reasonable since galaxies with high SFR tend to be massive according to the main-sequence of star-forming galaxies \citep[e.g.,][]{brinchmann04,daddi07}. 
Galaxies with the highest SFR at $z>0.8$ exhibit significantly strong galaxy clustering. 
The results of observational studies have shown that galaxies with high SFR reside in high-density environment  at $z\sim1$ \citep[e.g.,][]{cooper08,tran10}. 
In addition, the strong clustering signals support the idea that a high SFR at $z\sim1$ is mainly induced by galaxy mergers/interactions \citep[e.g.,][]{bekki01} in highly density regions, where galaxies are strongly clustered with each other. 

In contrast to high-SFR galaxies, the low-SFR end is almost constant and this trend could be caused by the mixture of galaxy populations in our samples. 
Red, passive galaxies are measured with strong clustering compared to blue, star-forming galaxies \citep[e.g.,][]{mccracken10,zehavi11,bielby14} and the correlation lengths of our low-SFR samples could be diluted by the mixture of strong clustering of red galaxies and the weak clustering of small blue galaxies. 
The discrepancy between our study and red galaxies of \citet{mostek13} can be explained based on the same reason. 

As previously indicated, the SFR is not an adequate indicator of halo masses due to the lack of an established halo occupation model and the absence of a constraint on the galaxy abundance. 
Therefore, we use only SM samples for the following discussions. 

\subsection{$\Mmin$ and $M_{1}$} \label{subsec:mmin-m1}
\begin{figure}[tpb]
\epsscale{1.25}
\plotone{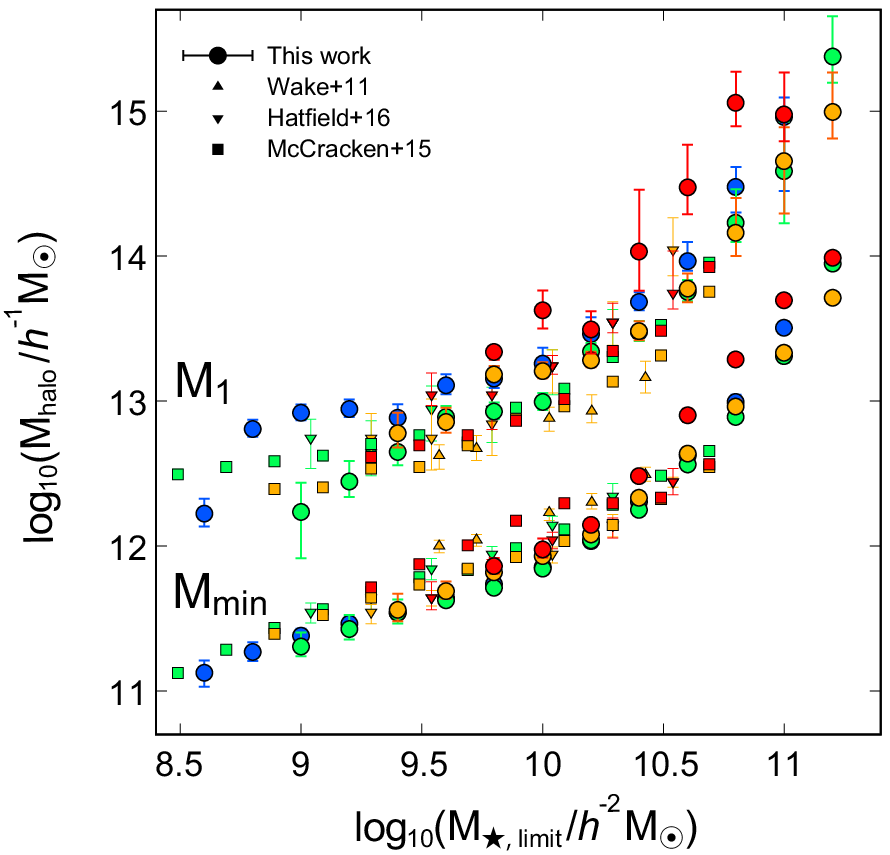}
\caption{The HOD mass parameters, $M_{1}$ (above symbols) and $\Mmin$ (bottom symbols), as a function of the stellar-mass limit. For comparison, we plot results of the photo-$z$ selected galaxies by \citet[][upward triangles]{wake11} at $0.9<z<1.3$ (orange), \citet[][downward triangles]{hatfield16} at $0.50<z<0.75$ (green), $0.75<z<1.00$ (orange), and $1.00<z<1.25$ (red), and \citet[][squares]{mccracken15} at $0.5<z<0.8$ (green), $0.8<z<1.1$ (orange), and $1.1<z<1.5$ (red), respectively. The units of the dark halo masses are unified in $h^{-1}\Msun$ in a logarithmic scale.}
\label{fig:Mmin-M1}
\end{figure}

We present $\Mmin$ and $M_{1}$ of our HSC galaxies as a function of stellar-mass limit in Figure~\ref{fig:Mmin-M1}. 
The results of previous HOD studies that adopt the same halo occupation function (equation~\ref{eq:hof}) and SM samples are also shown. 

The observed results show that both $\Mmin$ and $M_{1}$ monotonically increase with the stellar-mass limit, albeit with a small fluctuation in $M_{1}$. 
For $\Mmin$, this increasing trend indicates that the less-massive dark haloes can only possess less-massive centrals and vice versa. 
This is consistent with the trend of the stellar mass--halo mass relation of central galaxies that was derived using the abundance-matching technique \citep{behroozi10,behroozi13,moster10,moster13} as well as observational studies \citep{leauthaud12,coupon15}. 
The increasing trend of $M_{1}$ indicates that massive satellite galaxies form only in the massive dark haloes. 
This implies that a large fraction of satellite galaxies was originally massive central galaxies that became satellite galaxies through halo mergers. 

Considering the other observational results, $\Mmin$ shows little redshift evolution from $z = 1.5$ to $0.3$, indicating that central galaxies form similar condition in the underlying dark matter fluctuation at least at $z=0.3-1.5$. 
This result is consistent with the trend of small evolution of the stellar mass--halo mass relation of central galaxies from $z = 1.5$ to $z = 0.5$ with dark halo mass of $\Mh = 10^{11-13} \Msun$ presented by \citet{moster13}. 

$M_{1}$ also show an increasing trend with redshift, although there is large fluctuation compared to $\Mmin$. 
It can be determined that $M_{1}$ at $1.1 \leq z \leq 1.4$ (especially for massive galaxies) clearly shows the excess compared to galaxies at $z<1.1$, implying that satellite galaxies rarely form at $z \gtrsim 1$. 
The discussion of satellite fraction (Section~\ref{subsec:satellite_fraction}) considers the case of a small number of satellite galaxies at $z \gtrsim 1$. 

\subsection{Satellite Fraction} \label{subsec:satellite_fraction}
\begin{figure}[tpb]
\epsscale{1.25}
\plotone{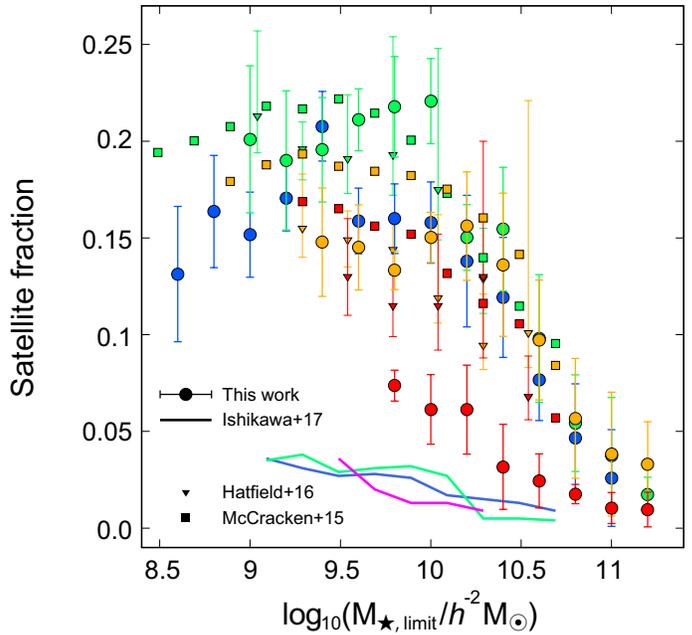}
\caption{Satellite fractions of each redshift bin as a function of the stellar-mass threshold. We show the satellite fractions of galaxies at $0.30 \leq z < 0.55$ ($z_{1}$; blue), $0.55 \leq z < 0.80$ ($z_{2}$; green), $0.80 \leq z < 1.10$ ($z_{3}$; orange), and $1.10 \leq z \leq 1.40$ ($z_{4}$; red), respectively. For comparison, we plot results of the photo-$z$ selected galaxies by \citet[][downward triangles]{hatfield16} at $0.50<z<0.75$ (green), $0.75<z<1.00$ (orange), and $1.0<z<1.25$ (red), and \citet[][squares]{mccracken15} at $0.5<z<0.8$ (green), $0.8<z<1.1$ (orange), and $1.1<z<1.5$ (red). Satellite fractions of $u$- (blue), $g$- (green), and $r$-dropout galaxies (magenta) calculated by \citet[][solid lines]{ishikawa17} are also shown to connect our results to higher-$z$ results. }
\label{fig:fsat}
\end{figure}

We calculate the satellite fraction of HSC galaxy samples following the equation \ref{eq:fs} as a function of stellar-mass limit using the best-fitting HOD parameters. 
The results are shown in Figure~\ref{fig:fsat} compared with results at similar redshift ranges from the literature. 
Overall, our results are in good agreement with the trends outline in the previous studies; i.e., satellite fractions of less-massive galaxies at $z \lesssim 1$ are almost constant. 
A rapid decrease of the satellite fraction can be found at the massive end and the decreasing trend is predominant for $z>1$ galaxies. 
In addition, satellite fraction decreases with redshift by fixing the stellar-mass threshold. 

The satellite fraction at $z\lesssim1$ remains almost constant up to $\logMstarlimit \sim$ 10.4, whereas it decreases drastically at the massive end. 
The steep decrease at the massive end can be caused by the efficient merging of massive infalling galaxies due to the short dynamical friction timescale \citep[e.g.,][]{colpi99,gan10}, and less-massive infalling galaxies can survive longer as satellite galaxies by orbiting their central galaxy. 
Moreover, steep decrease of the satellite fraction at the massive ends is also attributed to the small abundance of massive galaxies. 
According to the studies of stellar-mass function, abundance of $\Mstar\gtrsim10^{11}\Msun$ galaxies decrease drastically \citep[e.g.,][]{muzzin13b,tomczak14}; therefore, the number of galaxy pairs with  $\Mstar \gtrsim 10^{10.5}h^{-2}\Msun$ is thought to be quite small. 

The satellite fraction of our highest-$z$ bin ($1.10 \leq z \leq 1.40$) shows a significant drop compared to those at $z<1$ even for less-massive galaxies, which is also seen in the $z>3$ results using LBG samples \citep{ishikawa17,harikane18}. 
This may be due to the rarity of massive galaxies in high-$z$; less-massive galaxies live as central galaxies rather than satellite galaxies in the vicinity of massive galaxies, which are quite rare objects compared to the low-$z$ Universe. 
In addition to the small abundance of massive galaxies, efficient tidal disruption and high merger rate in high-$z$ can also be attributed to the small satellite fraction. 
\citet{wetzel11} found that the high-$z$ subhaloes show more radial orbits and sink deeper into the host haloes. 
It indicates that high-$z$ subhaloes can merge to their central with short time scale. 
In fact, observational studies as well as the numerical simulations have shown that the galaxy merger rate in $z>1$ is higher than $z<1$ \citep[e.g.,][]{conselice09,stott13,rodriguez-gomez15}. 
Moreover, the efficient tidal stripping due to the strong tidal force can reduce the stellar mass of high-$z$ satellites, ruling out less-massive high-$z$ satellites from observations due to the dimming their luminosity and/or the stellar-mass threshold. 
Therefore, the small satellite fraction in $z>1$ can be explained by the rarity of massive galaxies, high merger rate, and the strong tidal force of high-$z$ subhaloes. 

Contrary to our result, the satellite fraction of \citet{mccracken15} at $1.1<z<1.5$ does not show a significant drop, which is in disagreement with our results. 
This may be due to the different treatments of the HOD parameters of $\sigmaM$ and $\alpha$, which control the fraction of central galaxies in the vicinity of $\Mmin$ and the number of satellites in massive haloes. 
It is unclear whether these parameters were varied in the HOD-model fitting in \citet{mccracken15}. 
Even if they varied all the parameters, these two HOD parameters could not be strongly constrained by their limited angular scale $(\theta \lesssim 0^{\circ}_{\cdot}1)$, due to the small patch of the sky of the COSMOS. 

\subsection{Large-scale Galaxy Bias} \label{subsec:bias}
\begin{figure}[tpb]
\epsscale{1.25}
\plotone{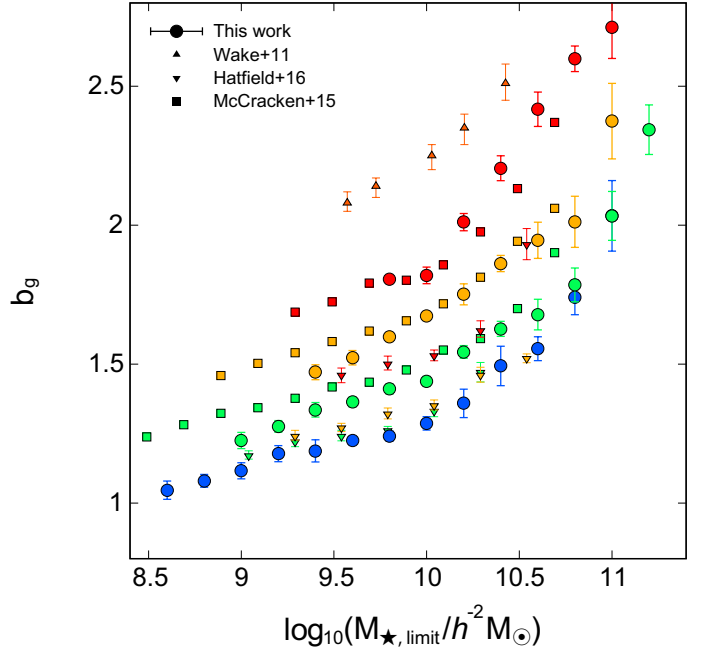}
\caption{The effective large-scale galaxy bias of each redshift bin as a function of the stellar-mass threshold. The galaxy biases of the HSC galaxy samples are $0.30 \leq z < 0.55$ ($z_{1}$; blue), $0.55 \leq z < 0.80$ ($z_{2}$; green), $0.80 \leq z < 1.10$ ($z_{3}$; orange), and $1.10 \leq z \leq 1.40$ ($z_{4}$; red), respectively. Observational results selected by the photometric redshifts are also plotted for the comparison: \citet[][upward triangles]{wake11} at $0.9<z<1.3$ (red), \citet[][downward triangles]{hatfield16} at $0.50<z<0.75$ (green), $0.75<z<1.00$ (orange), $1.0<z<1.25$ (red), and \citet[][squares]{mccracken15} are at $0.5<z<0.8$ (green), $0.8<z<1.1$ (orange), and $1.1<z<1.5$ (red), respectively. }
\label{fig:bg}
\end{figure}

The effective large-scale galaxy bias of each redshift is computed using the best-fitting HOD parameters and presented in Figure~\ref{fig:bg}. 
For comparison, the results obtained by recent large photometric surveys \citep{wake11,mccracken15,hatfield16} are also plotted. 
The trend of our results is consistent with referenced studies; galaxies are more biased by the stellar-mass limit and redshift. 
The galaxy bias of less-massive galaxies in low-$z$ is $\bg\sim1$, indicating that these galaxy samples are good tracers of invisible underlying dark matter distribution. 

The large-scale galaxy biases of the HSC galaxies are almost in good agreement with the results of \citet{mccracken15}. 
However, our results are significantly different from \citet{wake11} and \citet{hatfield16}; the large-scale galaxy biases of \citet{wake11} are significantly larger than ours, whereas those of \citet{hatfield16} are smaller than our results. 
One possible reason for the inconsistency with \citet{wake11} is due to the cosmic variance. 
The survey field of \citet{wake11} consists of two distinct fields; the COSMOS field \citep[$0.201$ deg$^{2}$,][]{scoville07} and the AEGIS field \citep[$0.189$ deg$^{2}$,][]{davis07}. 
Several studies have reported that the clustering strength in the COSMOS field at $z\sim1$ differs from that of other fields due to the cosmic variance \citep{mccracken08,meneux09,clowes13}. 
The ACF in the COSMOS field of \citet{wake11} at $0.9<z<1.3$ show stronger clustering at large scale compared to that of the AEGIS field. 
In addition, averaged ACFs were obtained by combining the ACFs observed in the both fields. 
Therefore, it is possible that the strong correlation in the large scale originated from the COSMOS-field-enhanced large-scale galaxy biases. 
It is noted that our galaxy biases only increased by a maximum of $10\%$ by using the same condition as \citet{wake11}, i.e., fixing the two HOD free parameters ($\sigmaM=0.15$ and $\alpha=1.0$) and adopting the cosmological parameters of the seven-year $WMAP$ results \citep{komatsu11}. 

The difference between our results and \citet{hatfield16} at $z>0.5$ can be attributed to the difference in the definition of the galaxy bias, although galaxy biases are comparable for the different definitions in the local Universe \citep{zehavi11}. 
In the studies highlighted in Figure~\ref{fig:bg} except for \citet{hatfield16}, the averaged large-scale galaxy biases were calculated using equation~\ref{eq:bgal}, whereas \citet{hatfield16} derived the galaxy bias based on its definition as equation~\ref{eq:bgal2}. 
It is reasonable for the consistency of $\Mmin$ among different studies that the difference of $\bg$ is due to its definition. 

\section{Discussion} \label{sec:discussion}
\subsection{Stellar-to-halo Mass Ratio} \label{subsec:shmr}
\begin{figure*}[tpb]
\epsscale{1.25}
\plotone{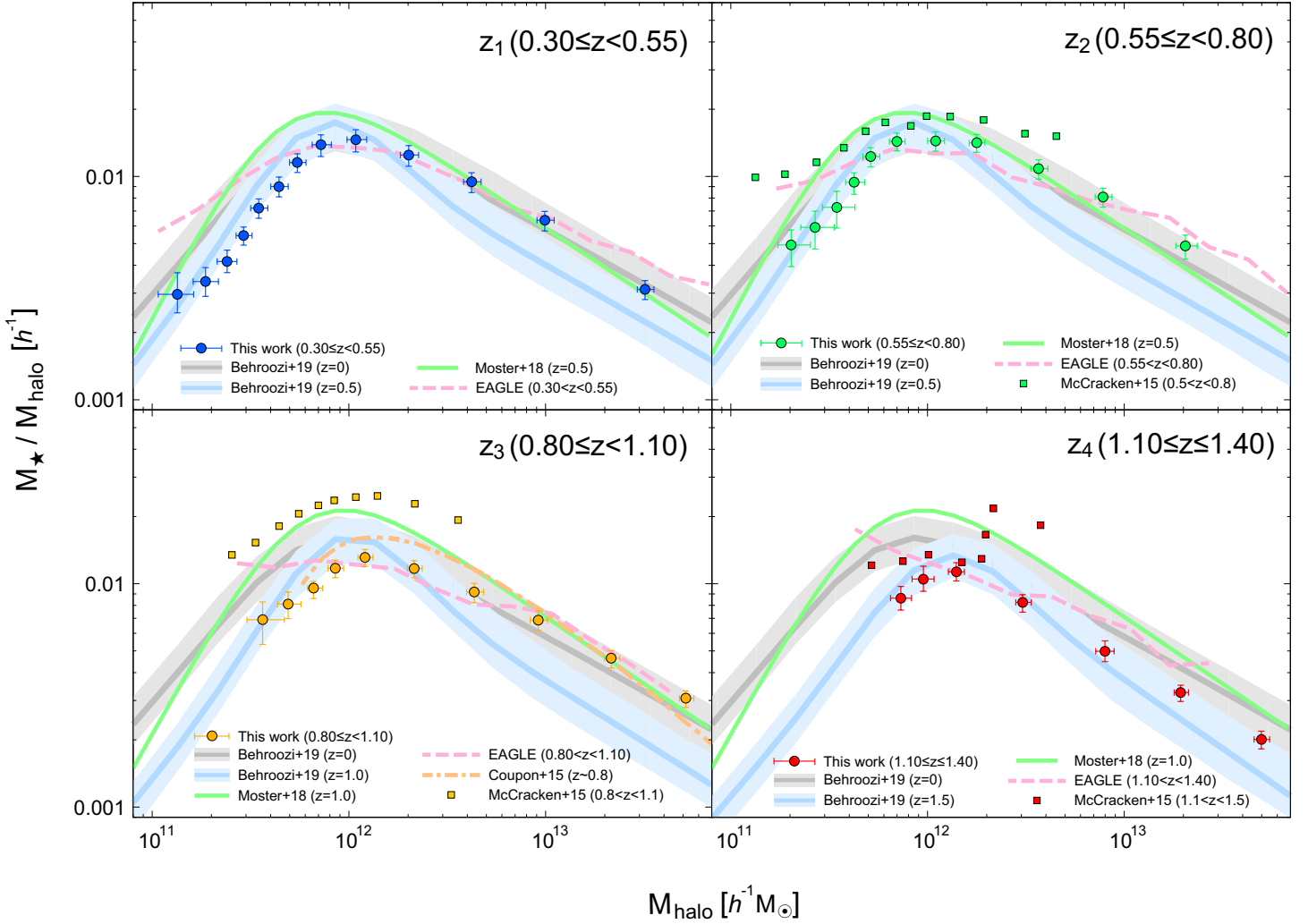}
\caption{Comparison of stellar-to-halo mass ratios (SHMRs). The SHMRs of the HSC SSP galaxies obtained in the redshift bin of $z_{1}$ ($0.30 \leq z < 0.55$; top left), $z_{2}$ ($0.55\leq z<0.80$; top right), $z_{3}$ ($0.80\leq z<1.10$; bottom left), and $z_{4}$ ($1.10 \leq z \leq 1.40$; bottom right) are shown. We also plot SHMRs of central galaxies computed by theoretical predictions using the \citet[][solid lines with shaded region]{behroozi18}, \citet[][solid lines]{moster18}, and the EAGLE simulations \citep[dashed lines;][]{schaye15,crain15,mcalpine16}, as well as observational results obtained by \citet[][dot-dash lines]{coupon15} and \citet[][squares]{mccracken15}. The SHMRs of \citet{mccracken15} are calculated using the stellar-mass limit and $\Mmin$ derived by their HOD-model analysis to match the definition of other studies. It is noted that the shaded regions of \citet{behroozi18} represent the $1\sigma$ confidence intervals. }
\label{fig:shmr}
\end{figure*}

Our stellar-to-halo mass ratios (SHMRs) based on the HSC SSP galaxy samples are presented in Figure~\ref{fig:shmr}. 
In the $\Mh$ versus $\Mstar/\Mh$ diagram, the stellar-mass limit, and the $\Mmin$ obtained using the HOD-model fitting analysis are used as the stellar mass and the dark halo mass of each stellar-mass bin, respectively. 
We also plot the theoretical predictions as well as the observational results: the latest results of the empirical models \citep{moster18,behroozi18}, results from the EAGLE simulations \citep{schaye15,crain15}, and the results from the deep- \citep{mccracken15} and the wide-field \citep{coupon15} extensively multi-wavelength photometric surveys. 
The SHMRs of the EAGLE simulations are calculated using central galaxies that satisfy the stellar-mass limits and the magnitude thresholds of this study. 
It should be noted that the definition of our SHMR (i.e., $SHMR(\Mh, z) = M_{\star, {\rm limit}}/\Mmin$) is the same as \citet{coupon15} and \citet{behroozi18}, and the SHMRs of \citet{mccracken15} and the EAGLE simulations are calculated by matching to our definition. 

Most strikingly, we succeeded in calculating the SHMRs that cover a wide halo mass range ($\Mh \sim 10^{11-14}h^{-1}\Msun$). 
The wide and relatively deep photometric data of the HSC Wide layer enable the capture of halo masses with peaked SHMR (pivot halo mass, $\Mpivot$) and both massive/less-massive slopes at $0.3 \leq z \leq 1.4$. 
Our SHMRs are generally consistent with referenced studies; the SHMRs have a peak at $\Mpivot \sim 10^{12}\Msun$, the peak values of SHMRs are $\sim 0.01-0.02$ regardless of its redshift, and the SHMRs gradually decrease toward the high and low-mass ends from its peak. 
However, our results differ slightly from \citet{behroozi18} in the slope of the massive end; i.e., our SHMRs show $1.5-2\sigma$ level excess compared to \citet{behroozi18} for $z \lesssim 1$ results. 
In this study, SHMRs are calculated using the massive galaxies selected by the HSC SSP Wide-layer over $\sim 145$ deg$^{2}$, in which the number of massive galaxies is large enough to apply HOD-model analysis and derive the halo mass precisely, and the accuracy of the SED fitting is thought to be reliable for massive, bright galaxies. 
The empirical model of \citet{behroozi18} used the stellar-mass function for the observational constraint and they adopted the stellar-mass function of \citet{moustakas13}, which is derived based on the observation over $5.5$ deg$^{2}$ in total. 
Therefore, the abundance of massive galaxies ($\Mstar \gtrsim 10^{11}\Msun$) could be underestimated in their study. 

In contrast to the massive end, our SHMRs at smaller masses than $\Mpivot$ haloes are consistent with previous studies, but show systematically smaller values compared to the model predictions, as well as other observational studies, albeit within the $1\sigma$ confidence intervals of \citet{behroozi18}. 
This difference can be caused by differences due to the of stellar-mass estimation method. 
The method of stellar-mass estimation is different among the studies, i.e., this study, \citet{mccracken15}, and \citet{coupon15} use the SED-fitting technique and \citet{behroozi18} and \citet{moster18} are constrained by the observational results such as the stellar-mass function. 
Even among the studies using the SED-fitting technique, stellar-mass estimation can be affected by difference in the physical assumptions such as the main sequence of star-forming galaxies, dust-extinction law, star-formation histories, and metallicity. 
The pivot halo masses and gradients of massive/less-massive end slopes of our results are almost comparable to \citet{mccracken15}; thus, the gap of the SHMR amplitude between these studies may be caused by differences in the physical assumptions. 

\subsection{Connection to High-$z$ SHMRs} \label{subsec:connect_to_high-z_shmr}
\begin{figure}[tpb]
\epsscale{1.25}
\plotone{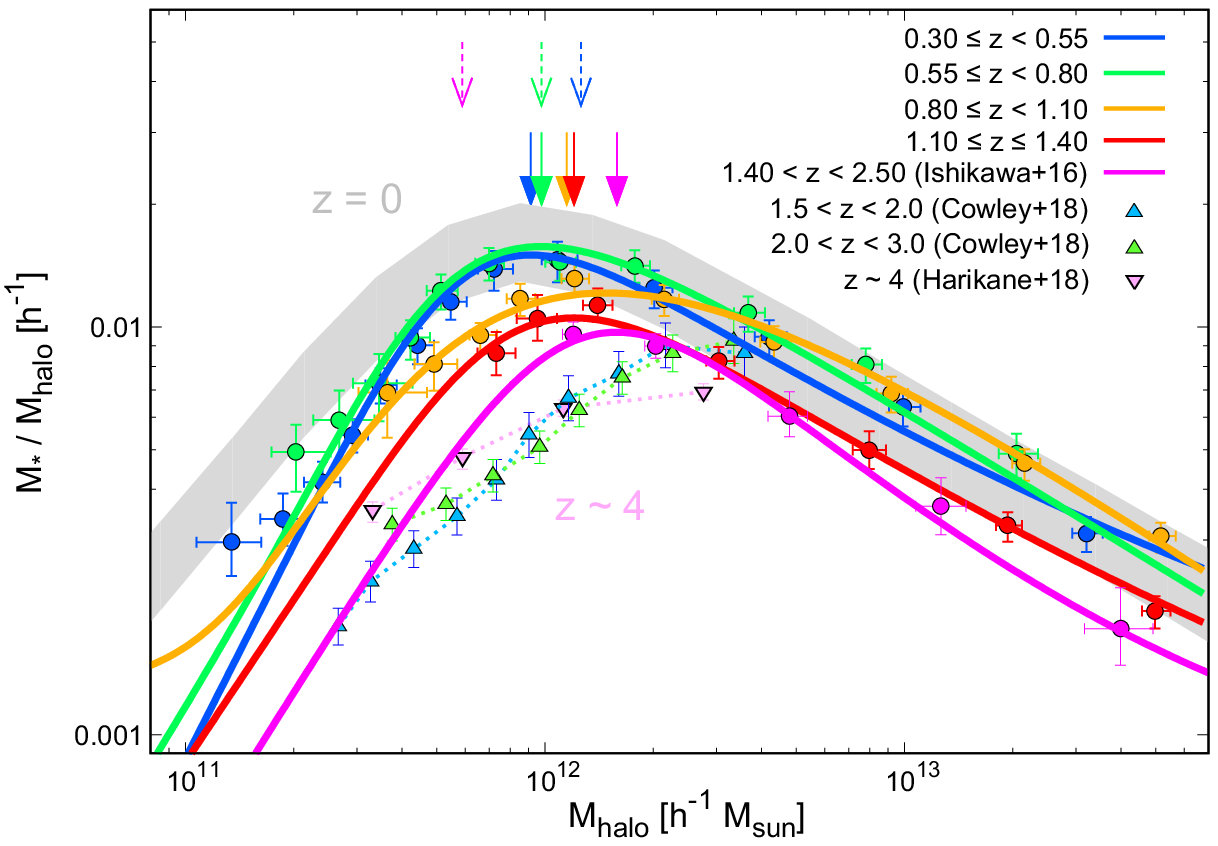}
\caption{Redshift evolution of observed SHMRs. Our observational results are plotted as blue ($0.30 \leq z <0.55$), green ($0.55 \leq z <0.80$), orange ($0.80\leq z <1.10$), and red ($1.10 \leq z <1.40$) circles, respectively. The solid lines are the best-fitting SHMRs of each redshift bin fitted using the parameterized relation presented by \citet{behroozi13} and the arrows above our SHMRs represent the $\Mpivot$ of the fitted SHMRs. The dashed arrows on top of the panel also show the $\Mpivot$ of $z\sim3$ (blue), $z\sim4$ (green), and $z\sim5$ (red) dropout galaxies calculated by \citet{ishikawa17}. The gray shaded region indicates the model prediction of SHMR at $z=0$ with $1\sigma$ confidence level by \citet{behroozi18}. To connect the SHMRs of the HSC SSP to higher-$z$, we also show the SHMRs using star-forming galaxies at $1.40<z<2.50$ \citep[magenta circles][]{ishikawa16}, the photometric-redshift selected galaxies at $1.5<z<2.0$ (blue triangles) and $2.0<z<3.0$ (green triangles) by \citet{cowley18}, and dropout galaxies at $z\sim4$ (red downward triangles) by \citet{harikane18}. }
\label{fig:z-evo_shmr}
\end{figure}

We connect our SHMRs observed in $0.3\leq z \leq 1.4$ to higher-$z$ results to trace the redshift evolution of the stellar-mass assembly of galaxies. 
To clearly show the characteristics of SHMRs, we here use the formulated SHMR model. 
We employ the parameterized relation of the SHMR presented by \citet{behroozi13}. 
The best-fitting models for our observed SHMRs of each redshift bin are presented in Figure~\ref{fig:z-evo_shmr}, and the pivot halo masses and values of peak SHMR in the best-fitting models are shown in Table~\ref{tab:shmr_fit}. 
To connect our results to those beyond $z=1.4$, we also plot the observed SHMRs at $z\sim2$ \citep[sgzK galaxies][]{ishikawa16}, $z=1.5-2.0$ and $z=2.0-3.0$ \citep{cowley18}, and $z\sim4$ \citep{harikane18}. 
It should be noted that these high-$z$ galaxy populations are not necessarily the same population as we observed at $z<1.4$. 

The SHMRs at $z<0.8$ rarely evolve from the model prediction at $z=0$; both $\Mpivot$ and peak amplitude exhibit little change. 
In comparison, the SHMRs at $z>1$ show an apparent evolution from $z=0$; $\Mpivot$ clearly shifts towards a higher halo mass and the peak amplitude decreases with the redshift, which is consistent with a model prediction presented by \citet{rodriguea-puebla17}. 
The difference in evolutional trend given that $z\sim1$ can be interpreted to be attributed to the evolution of stellar-mass function. 
Observational studies have revealed that the evolution of the stellar-mass function is quite small over $0.2<z<1.0$ and amplitude of the Schechter function decreases with redshift at $z>1$ \citep[e.g.,][]{muzzin13b,tomczak14}. 
The decrease of the peak value of SHMR at $z>1$ can be a result of the decrease of the total amount of stellar components in the high-$z$ Universe. 

An increase in $\Mpivot$ with redshift, especially at $z>1$, indicates that the supernova feedback mechanism for the suppression of the stellar-mass assembly in low-mass dark haloes is more efficient than $z\sim0$. 
The quenched fraction of central galaxies with stellar mass $\Mstar \sim 10^{10}\Msun$, which is typically hosted by dark haloes with $\Mpivot$, evolves from $\sim0.4$ at $z\sim0$ into $\sim0.1$ at $z\sim1$; therefore, most of the galaxies at $z>1$ are star-forming galaxies, even in the mass range in which quiescent galaxies are dominated in the local Universe \citep[e.g.,][]{drory09,tinker13}. 
Active star-forming galaxies occupy relatively massive haloes at $z>1$ and $\Mpivot$ gradually shifts to lower-mass haloes by increasing the fraction of quiescent galaxies in $\Mpivot$ haloes towards $z\sim0$. 

Qualitatively, the peak amplitudes of SHMRs in $z>2$ are smaller than the results for $z<2$, suggesting that star formation is still in progress at high-$z$. 
The low-mass slopes of \citet{cowley18} that are derived using unbiased photo-$z$ samples similar to this study, exhibit a shallower  gradient, implying that star-formation activity in high-$z$ galaxies hosted by $\Mh \lesssim \Mpivot$ haloes are highly suppressed by the supernova feedback effect. 
The SHMR of \citet{harikane18} also pronouncedly shows the same trend; however, it should be noted that their galaxy sample is consist only of star-forming galaxies. 

\begin{deluxetable}{lcc}
\tablecaption{Best-fitting Values of the Pivot Halo Mass and the SHMRs at Pivot Halo Mass}
\tablehead{Redshift & $\log_{10}(\Mpivot/h^{-1}\Msun)$ & $\Mstar/\Mpivot (\times 10^{-2}h^{-1})$}
\startdata
(HSC SSP) &  &  \\
$0.30 \leq z < 0.55$ & $11.96 \pm 0.02$ & $1.501 \pm 0.086$ \\
$0.55 \leq z < 0.80$ & $11.99 \pm 0.01$ & $1.574 \pm 0.067$ \\
$0.80 \leq z < 1.10$ & $12.06 \pm 0.02$ & $1.201 \pm 0.088$ \\
$1.10 \leq z \leq 1.40$ &12.08 $ \pm 0.04$ & $1.052 \pm 0.040$ \\
\hline
(sgzK galaxies) & & \\
$1.40 \lesssim z \lesssim 2.50$ & $12.20 \pm 0.04$ & $0.970 \pm 0.044$ \\
\enddata
\tablecomments{The SHMRs of sgzKs are obtained by resampling the sgzKs of \citet{ishikawa15,ishikawa16} according to its stellar mass and by reperforming the HOD-model analysis on them again.}
\label{tab:shmr_fit}
\end{deluxetable}

\subsection{Evolution of Pivot Halo Mass} \label{subsec:evolution_Mpivot}
\begin{figure*}[tpb]
\plotone{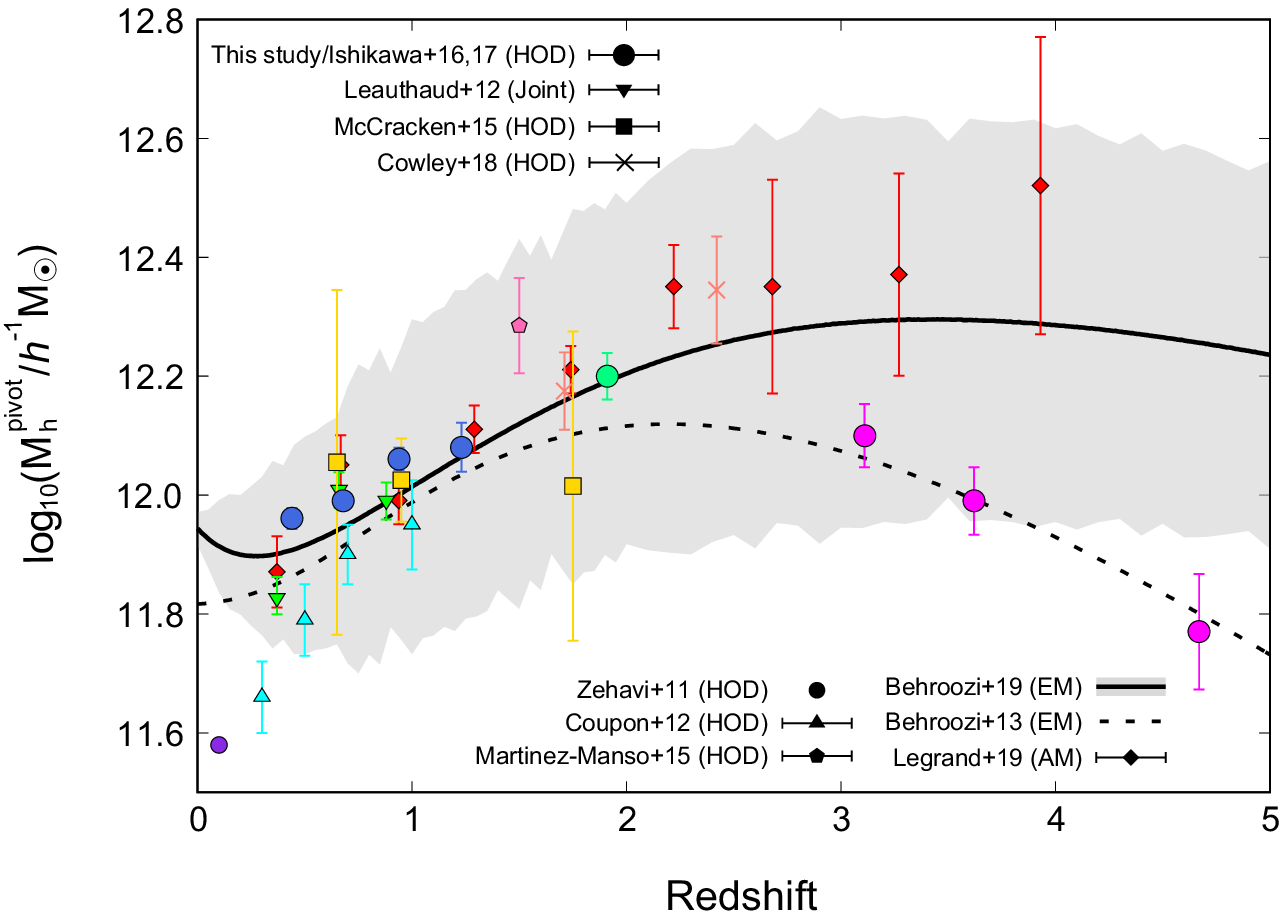}
\caption{The redshift evolution of $\Mpivot$. The blue circles represent our results obtained in the HSC SSP Wide layer, and the green circle and the magenta circles are the results of sgzK galaxies \citep{ishikawa16} and Lyman break galaxies \citep{ishikawa17}, respectively. We also plot observational results at various redshifts: stellar-mass limited samples that are similar to this study presented by \citet[][green downward triangles]{leauthaud12}, \citet[][orange squares]{mccracken15}, and \citet[][red crosses]{cowley18}, as well as luminosity-limited samples presented by \citet[][purple circle]{zehavi11}, \citet[][cyan upward triangles]{coupon12}, and \citet[][pink pentagon]{martinez-manso15}. It is noted that  the joint-analysis technique is only used in \citet{leauthaud12} and other observational studies adopt the HOD-model analysis. The red diamonds represent the results of the SHAM in the COSMOS field presented by \citet{legrand18}. The model-predicted redshift evolution of $\Mpivot$ are also shown: the dotted line and the solid line are the predictions of the empirical models by \citet{behroozi13} and \citet{behroozi18}, respectively. The gray shaded region represents the $1\sigma$ confidence level of \citet{behroozi18} obtained by varying all the parameters of the formulated SHMR relation within $1\sigma$ intervals. The characters listed after the names of the authors denote the analysis method used in each study. }
\label{fig:Mpivot}
\end{figure*}

The redshift evolution of our $\Mpivot$ is presented in Figure~\ref{fig:Mpivot}. 
For comparison, we also plot other observational results for various redshifts as well as theoretical predictions using the AM technique. 
The observed results are obtained using the HOD-model analyses based on the luminosity-limited samples \citep{zehavi11,coupon12,martinez-manso15} and the SM samples \citep{leauthaud12,mccracken15,ishikawa16,ishikawa17,cowley18}, respectively. 
In addition to the aforementioned clustering studies, the results obtained via the empirical model \citep{behroozi13,behroozi18} and the subhalo AM technique \citep[SHAM;][]{legrand18} are also shown. 

The $\Mpivot$ in the HSC SSP are consistent with previous studies based on the SM samples. 
The most efficient star-formation activity is in progress within dark haloes with masses of $\Mh \sim 10^{12}h^{-1}\Msun$ up to $z\sim1$, and $\Mpivot$ gradually increases with the redshift. 
Although there is a little evolution ($\sim \pm 0.2$ dex) in the pivot halo mass, we systematically show that the pivot halo mass is almost constant $(\Mpivot \sim 10^{12}\Msun)$ up to $z\sim5$, using only clustering and HOD-model analyses. 
The result suggests that most of the star formation occurs in a narrow range of halo masses, which is consistent with the theoretical predictions for galaxy formation models \citep[e.g.,][]{ree77,wang13,dekel19}. 

By tracing the redshift evolution, it is shown that $\Mpivot$ monotonically increases with redshift, at least up to $z\sim2$. 
Beyond $z\sim2$, there are two scenarios for $\Mpivot$ evolution: $\Mpivot$ increases monotonically \citep{cowley18,legrand18}, otherwise it decreases with redshift \citep{behroozi13,ishikawa17}. 
The latest result of the model prediction by \citet{behroozi18} is in the middle of these scenarios; $\Mpivot$ increase monotonically up to $z\sim3$ and then gradually decreases with redshift. 
However, theoretical approach contains relatively large uncertainties in high-$z$. 
Observational studies suggest unbiased photo-$z$ selected galaxies show increasing evolution \citep{cowley18,legrand18}, whereas $\Mpivot$ of star-forming dropout galaxies decrease with redshift \citep{ishikawa17}. 
To understand the star formation in high-redshift dark haloes, it is necessary to measure $\Mpivot$ in $z\gtrsim2$ with higher statistical precision and various galaxy populations. 

\subsection{Baryon Conversion Efficiency} \label{subsec:bce}
We calculate the instantaneous baryon conversion efficiency (BCE) of our galaxy samples. 
The BCE, $\epsilon\left(\Mh, z\right)$, is defined as the mass ratio between accreted baryons and formed stellar components as, 
\begin{equation}
\epsilon\left(\Mh, z\right) = \frac{d\Mstar}{dt}/\frac{dM_{{\rm b}}}{dt}, 
\label{eq:bce}
\end{equation}
where $d\Mstar/dt$ and $dM_{{\rm b}}/dt$ represent the SFR and a baryon accretion rate (BAR), respectively. 
The BAR can be calculated as follows: 
\begin{equation}
{\rm BAR} = f_{{\rm b}} \times \frac{d\Mh}{dt},
\label{eq:bar}
\end{equation}
where $f_{{\rm b}}$ and $d\Mh/dt$ are a baryon fraction and a mass-growth rate of dark haloes, respectively. 
In our cosmological parameters, the baryon fraction is $f_{{\rm b}} = \Omega_{{\rm b}}/\Omega_{{\rm m}} \sim 0.159$. 
We employ the dark halo mass-growth rate as a function of the redshift and dark halo mass presented by \citet{fakhouri10}, which is based on the results of the $N$-body simulations. 

\begin{figure*}[tpb]
\epsscale{1.2}
\plotone{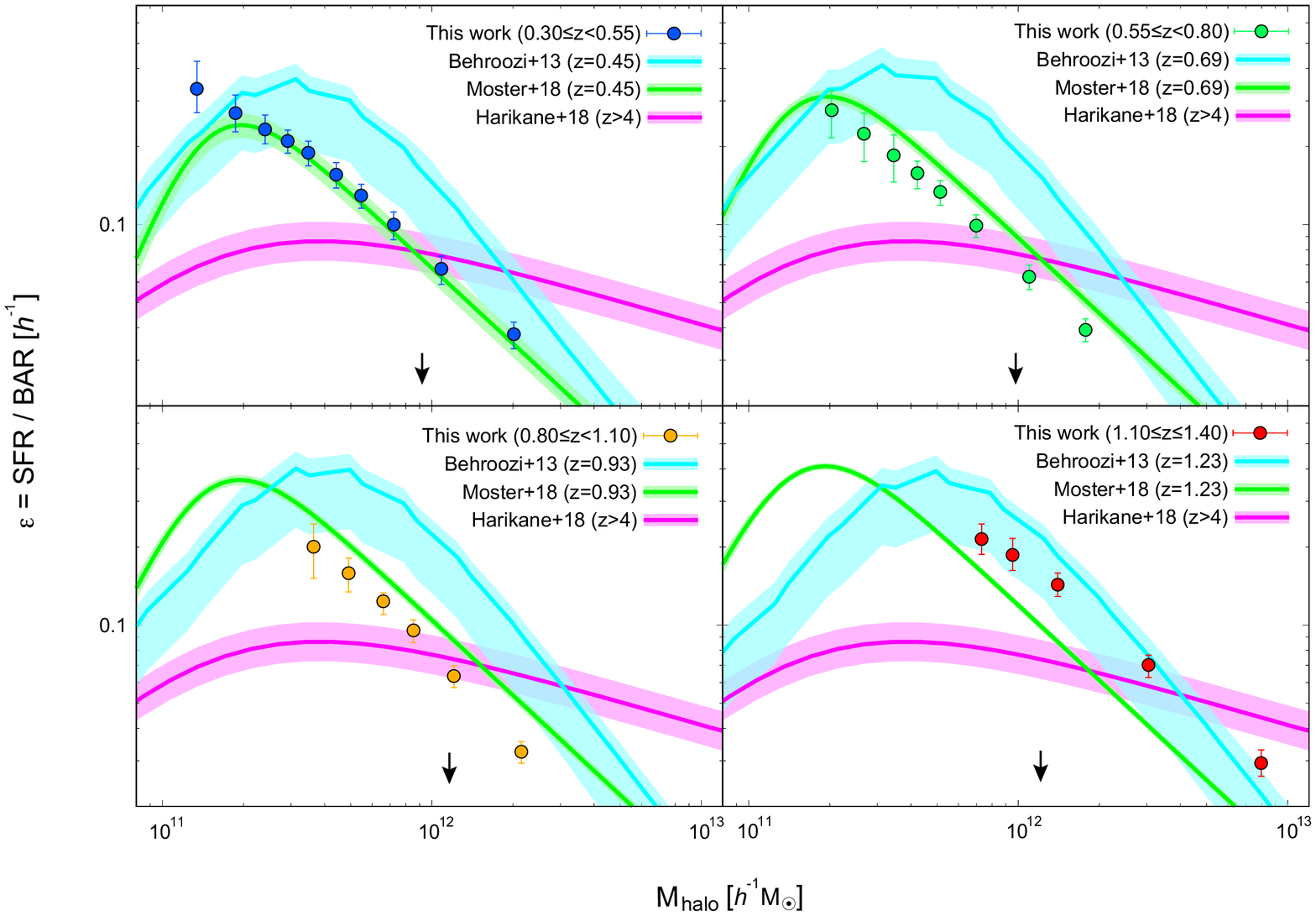}
\caption{Instantaneous baryon conversion efficiencies (BCEs) of the HSC galaxies at $0.30 \leq z < 0.55$ (top left), $0.55 \leq z < 0.80$ (top right), $0.80 \leq z < 1.10$ (bottom left), and $1.10 \leq z \leq 1.40$ (bottom right). We also plot the results of the empirical models presented by \citet[][cyan]{behroozi13} and the \citet[][green]{moster18} at the effective redshift of each redshift bin of this study, as well as the observational result using the Lyman break galaxies at $z>4$ given by \citet[][magenta]{harikane18}. The shaded regions of \citet{behroozi13} and the \citet{moster18} represent the $1\sigma$ confidence intervals, whereas those of the \citet{harikane18} are $0.15$ dex observational scatters, respectively. The result of \citet{harikane18} is plotted by dividing their SFR$/\dot{M_{{\rm h}}}$ versus $\Mh$ relation by the baryon fraction to match the definition. In calculating our BCEs, we use the results of the HOD analysis for the SM samples and BCEs are derived by dividing the averaged SFR by the baryon accretion rate (BAR) calculated using the $\Mmin$. The arrows above the $x$-axis represent the $\Mpivot$ in each redshift bin.}
\label{fig:bce}
\end{figure*}

The BCEs of the SM samples are presented in Figure~\ref{fig:bce}. 
Our BCEs are calculated using the BARs derived from $\Mmin$ and the averaged SFRs of each stellar-mass bin. 
In Figure~\ref{fig:bce}, we also show BCEs calculated using the empirical models of \citet{behroozi13} and \citet{moster18}, as well as the observational result in $z>4$  \citep{harikane18}. 
Our results are qualitatively consistent with the empirical models; the BCEs decrease with halo mass and the power-law slopes in massive ends are almost consistent with the models. 
For $0.30\leq z \leq 0.55$, the BCE of our lowest stellar-mass bin shows the highest value, whereas the BCEs of both empirical models turn to decrease at the corresponding halo mass. 

It is notable that the halo masses with the highest BCE are much smaller than the $\Mpivot$ of the SHMRs for each redshift, although the peak halo masses are slightly different among the studies. 
This mass gap between the peak halo masses of the BCE and the SHMR (i.e., $\Mpivot$) has already been shown using the empirical model at $z=0$ \citep{behroozi13b}, but we show that the mass gap can be seen at least up to $z=1.4$ by observation. 
Although accurate peak halo masses cannot be captured due to the lack of less-massive ends, the halo mass with the peak BCE seems to be $10^{11.0-11.5}h^{-1}\Msun$, and $\Mpivot$ in the SHMR diagram is $\sim10^{12}h^{-1}\Msun$, regardless of its redshift. 
Therefore, this implies that the most efficient star-forming activity is in progress in less-massive galaxies hosted by $10^{11.0-11.5}h^{-1}\Msun$ dark haloes. 
Considering the amplitude of the SHMR, galaxies hosted in $10^{11.0-11.5}h^{-1}\Msun$ dark haloes rapidly assemble stellar components much faster than the growth of dark halo mass. 
Moreover, they evolve towards galaxies hosted by $\Mpivot$ dark haloes in which the mass ratio of total stellar masses to dark haloes reaches its peak. 
Hence, $\Mpivot$ is the halo mass where galaxies have sufficiently evolved their stellar components compared to the growth of dark haloes, rather than the halo mass where galaxies can form stars most efficiently at that epoch. 

\begin{figure}[tpb]
\epsscale{1.25}
\plotone{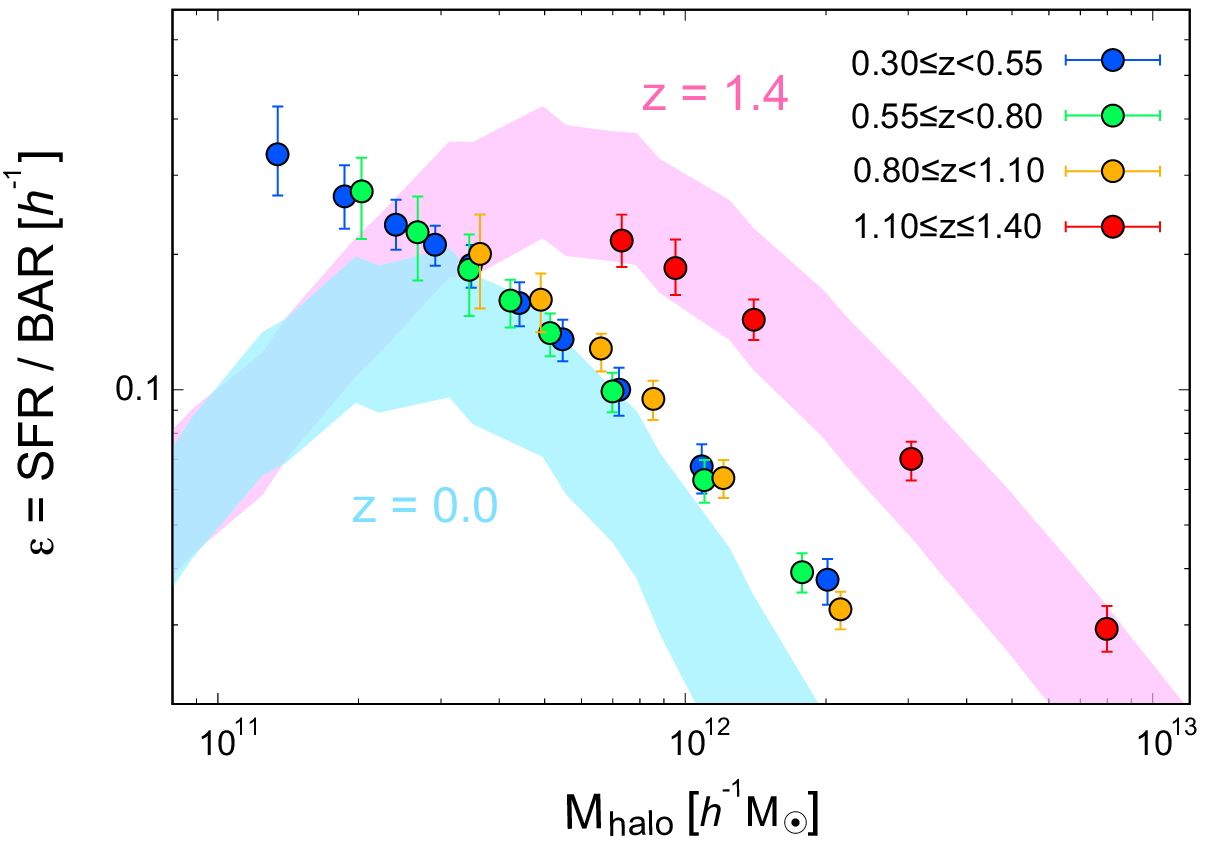}
\caption{Redshift evolution of our observed BCEs. The BCEs of HSC SM sample are plotted as blue ($0.30\leq z < 0.55$), green ($0.55\leq z < 0.80$), orange ($0.80\leq z < 1.10$), and red ($1.10\leq z \leq 1.40$). For comparison, the BCEs computed by the empirical model of \citet{behroozi13} at $z=0$ (blue shaded region) and $z=1.4$ (red shaded region) are also shown with $1\sigma$ uncertainties. }
\label{fig:bce_z-evo}
\end{figure}

Figure~\ref{fig:bce_z-evo} shows the redshift evolution of our BCEs. 
It is determined that the BCEs are almost constant up to $z\sim1$, whereas they show an apparent excess only at $z>1$. 
Dark matter only $N$-body simulations have shown that the BAR is much higher at high-$z$ \citep[e.g.,][]{fakhouri10,faucher11}; therefore, the absence of evolution of BCEs at $z<1$ suggests that, by fixing the halo mass, the star-formation rate of galaxies at $z<1$ decreases as fast as the decreasing rate of accreting baryonic matters. 
In contrast, $z\sim1.4$ is near the era in which the cosmic star-formation rate density is at a peak \citep[e.g.,][]{hopkins00,madau14} and star formation in dark halo is thought to be still efficient. 
This causes the excess of BCEs in $1.1\leq z \leq 1.4$. 
However, we should confirm whether the excess of BCEs occur only for $z>1$ by further observations. 

\subsection{Relationship Between Galaxies and Dark Haloes at $0.3 \leq \lowercase{z} \leq 1.4$}\label{subsec:discussion}
In the previous sections, we investigated the dark halo properties in the framework of the halo model. 
We will discuss several implications from this study, which focus on the galaxy evolution and its relation with their host dark haloes at $0.3 \leq z \leq 1.4$ along with other studies. 

\subsubsection{Effect of merging, {\it in situ}/{\it ex situ} star formation, and AGN feedback on central and satellite galaxy formation}\label{subsubsec:centra_and_satellite}
In our HOD analysis, the power-law slope in the $M_{\star, {\rm limit}}$ versus $\Mmin$ relation changes at $M_{\star, {\rm limit}}\sim 10^{10.5}h^{-2}\Msun$ (Figure~\ref{fig:Mmin-M1}). 
Fitted to our results, dark halo masses of central galaxies evolve according to $\Mmin \propto M_{\star, {\rm limit}}^{0.5}$ for $M_{\star, {\rm limit}} < 10^{10.5}h^{-2}\Msun$ galaxies and $\Mmin \propto M_{\star, {\rm limit}}^{1.7}$ for $> 10^{10.5}h^{-2}\Msun$ galaxies, respectively. 
This two-phase evolutionary trend can be understood as different growth modes of stellar mass: star-formation activity within galaxies ({\it in situ}) and galaxy mergers ({\it ex situ}). 
According to the results of numerical simulations, \citet{lackner12} reported that approximately only $15\%$ of the final stellar mass of galaxies with stellar masses of $\Mstar \sim 10^{10}\Msun$ at $z\sim0$ was acquired via galaxy merging (i.e., {\it ex situ} evolution), and \citet{rodriguez-gomez16} showed that the {\it ex situ} stellar-mass fraction rapidly increase for $\Mstar \gtrsim 10^{11}\Msun$ galaxies and
more than $90\%$ of stellar masses of $\Mstar<10^{10.5}\Msun$ are in situ stars, at least up to $z\sim2$. 
Therefore, the mass-growth rate between dark haloes and baryons ($\Mmin/M_{\star, {\rm limit}}$) substantially increase by changing the origin of dominated stellar masses whose formation sites are internal/external environments, which can be seen in the power-law slope of our observed $\Mmin$ versus the $M_{\star, {\rm limit}}$ diagram (Figure~\ref{fig:Mmin-M1}). 

It is worth noting that the redshift and stellar-mass dependence of the merger rate can quantitatively be constrained if the power-law slope at the massive end can be accurately measured in future observations. 
The power-law slope increases as the major merger rate increases because the dark halo mass effectively increases through major merger events, and the massive end of the $\Mmin$ versus $M_{\star, {\rm limit}}$ relation exceeds the $\Mmin \propto M_{\star, {\rm limit}}^{1.7}$ power-law relation if massive galaxies frequently experience major mergers compared to less-massive galaxies. 
In this study, we find that the $\Mmin \propto M_{\star, {\rm limit}}^{1.7}$ relation is almost unchanged up to $z=1.4$, indicating that the major merger rate does not largely depend on the stellar mass. 

In contrast to the case of central galaxies, the evolution of $M_{1}$ and the resultant satellite fraction is complicated (Figure~\ref{fig:Mmin-M1} and \ref{fig:fsat}). 
Why is the abundance of satellite galaxies at $z>1$ extremely small? 
As discussed in Section~\ref{subsec:satellite_fraction}, the possible reasons for the deficit high-$z$ satellite galaxies based on HOD analysis may be the lack of the galaxy pairs with comparable stellar masses. 
For instance, \citet{marmol-queralto12} found that nearly $30\%$ of massive galaxies at $z\sim2$ form galaxy pairs with a $1:100$ stellar-mass ratio. 
Therefore, the large number of satellite galaxies at $z>1$ cannot be attributed to satellite fraction for our stellar-mass range ($9.8\leq \logMstarh \leq 11.2$ for $z>1$). 
The HOD analysis with a much wider stellar-mass range at $z>1$ may resolve the discrepancy of satellite fraction at $z\sim1$. 

The small stellar masses of $z>1$ satellites could be attributed to the AGN feedback in addition to the efficient disruption due to the tidal stripping as discussed in Section~\ref{subsec:satellite_fraction}. 
\citet{dashyan19} investigated the effect of the feedback by the central AGNs on star formation of satellite galaxies based on cosmological zoom-in simulations and found that the AGN feedback significantly suppresses the star-forming activities of satellite galaxies at $z\sim2$ at the earliest. 
Observational studies have shown that the AGN activity has a peak at $z\sim2$ during the cosmic time, and the number density of the brightest AGNs steeply decreases after that epoch \citep[e.g.,][]{fan01,croom09}. 
Suppression of satellite galaxy formation will be reduced with the extinguishing of the AGNs, especially for luminous AGNs that have a large impact on the AGN feedback effect, and relatively massive satellites can start to evolve at $z<1$. 
Therefore, it is very meaningful for satellite galaxy formation to understand the AGN feedback effect quantitatively from the both aspects of the mass and its environment. 

\subsubsection{Downsizing of star formation at $0.3\leq z \leq 1.4$}\label{subsubsec:shmr_and_bce}
The small increase of $\Mpivot$ over $0.3\leq z \leq1.4$ (Figure~\ref{fig:Mpivot}) is largely related to the redshift evolution of the BCEs (Figure~\ref{fig:bce_z-evo}) in the framework of the galaxy downsizing \citep[e.g.,][]{cowie96,fontanot09}. 
For a fixed dark halo mass, our observed BCEs at $z>1$ are much higher than those at $z<1$. 
This is indicative of the anti-hierarchical formation and evolution scenario of galaxies, wherein massive galaxies have almost completed their stellar-mass assembly at $z>1$ \citep[e.g.,][]{bower06,collins09}. 
We will discuss the downsizing of star formation from the aspect of the redshift evolution of the halo mass of the peak BCE and $\Mpivot$. 

There is a gap between peaks of the BCE and $\Mpivot$ but it gradually decreases with redshift (Figure~\ref{fig:bce}). 
The redshift evolution of this gap can be explained by the redshift evolution of the cosmic star formation. 
\citet{juneau05} investigated the star-formation history over $0.8<z<2.0$ and found that the massive galaxies ($\Mstar>10^{10.8}\Msun$) are almost completed their star formation by $z\sim 1.5-2.0$, while the intermediate-mass galaxies ($\Mstar=10^{10.2-10.8}\Msun$) have a star-burst phase at $z\sim1$ and then they are in the quiescent evolution mode. 
Therefore, the slightly difference of the peak halo mass between the BCE and the SHMR in our highest-$z$ bin is indicative of the difference of this star-formation history between the most massive and the intermediate-mass galaxies. 

On the other hand, peaks of the BCE largely decrease with cosmic time, whereas the $\Mpivot$ show small decrease. 
At $z<1$, the intermediate-mass galaxies are in the passive evolution \citep{juneau05}, although less-massive galaxies continue to form stars \citep[e.g.,][]{ilbert10}, indicating that the cosmic star-formation shifts toward less-massive galaxies since $z\sim1$. 
According to studies of the star-formation history of dwarf galaxies, massive dwarf galaxies with $\Mstar \sim 10^{8-9}\Msun$, whose host haloes correspond to masses with the peak BCE, have rapidly assembled their stellar components at $0<z<0.5$ \citep{garrison-kimmel19}, which is consistent with the BCE peak in our study at $0.30\leq z \leq 0.55$. 
Consequently, the peak of the BCE reflects the dark halo mass with the instantaneously high star-forming activities at each epoch and the evident decrease of the peak BCE supports the downsizing of star formation. 
However, the SHMR represents the integrated star formation within dark haloes. 
The $\Mpivot$ may be roughly determined by the high star-forming activities of massive galaxies at $z\gtrsim2$ and the actively star formation of less-massive galaxies at $z<1$ cannot largely shift $\Mpivot$ since the cosmic star-formation rate density is small \citep[e.g.,][]{madau14}. 

\section{Summary} \label{sec:summary}
In this paper, we present the relationship between the baryonic properties of galaxies and their host dark haloes via the precision clustering and HOD analyses using a large amount of galaxy samples selected via the HSC SSP S16A data. 
Using the galaxy samples observed in the HSC SSP Wide layer, we obtain $\sim 5,000,000$ galaxies over $\sim 145$ deg$^{2}$ survey field down to $i=25.9$, which is a sufficient deep and wide data to investigate the evolutionary history of galaxies at $0.30 \leq z \leq 1.40$ and stellar-mass range ($8.6 \leq \logMstarh \leq 11.2$) compared to other observational studies. 
The SED-fitting technique evaluates fundamental physical parameters of baryons, i.e., galaxy stellar mass, star-formation rate (SFR), and photometric redshift, of each galaxy at $0.3 \leq z \leq 1.4$. 
We divide our galaxy samples into four distinct redshift bins, $0.30 \leq z < 0.55$, $0.55 \leq z < 0.80$, $0.80 \leq z < 1.10$, and $1.10 \leq z \leq 1.40$, according to their photometric redshifts. 
Furthermore, galaxy samples of each redshift bin are also divided into subsamples according to its stellar mass and SFR to reveal the redshift evolution and baryonic characteristics dependence of the galaxy--dark halo co-evolution. 

We measure the angular two-point auto correlation functions (ACFs) of our galaxy samples. 
Due to the large number of galaxy samples and the wide survey field of the HSC SSP Wide layer, the ACFs can be computed with a high S/N ratio even for a large-angular scale ($\sim 1^{\circ}$). 
Moreover, ACFs of rare objects such as massive galaxies ($\logMstarh \geq 11.2$) and highly star-forming galaxies ($\logSFR \geq 1.0$) can also be determined well. 
The HOD-model analysis is applied to our ACFs and the HOD framework obviously succeeds in representing observed galaxy clustering for all stellar-mass/SFR ranges and dynamical scales at $0.3 \leq z \leq 1.4$. 

Our major findings based on the halo-model analyses in this study can be summarized as follows: 
\begin{enumerate}

\item For SM samples, the correlation lengths show a monotonic increase with the stellar-mass limit, irrespective of its redshift, and steeply increase at $\logMstarlimit \gtrsim 10.4$ for each redshift bin. 
This suggests that massive galaxies a*re hosted by more massive dark haloes for each redshift. 
However, for SFR samples correlation lengths are almost constant up to $\logSFRlimit \sim 0$ and then increase steeply with the SFR limit. 
In low-SFR end, the galaxy clustering signal is thought to be diluted by the mixture of the strong correlation of red, passive galaxies and the weak correlation of low-SFR blue galaxies. 
In contrast, galaxies with highly SFR show a strong correlation regardless of their redshift. 
We can interpret such strong clustering signals as high SFR galaxies tend to reside in the over-density regions, where galaxies strongly cluster with each other, if high SFRs are mainly induced by galaxy interactions and/or mergers. 

\item Satellite fractions of HSC galaxies are almost consistent with previous observational studies. 
The satellite fractions of less-massive galaxies at $z\lesssim1$ are almost constant at $\sim20\%$, and they gradually decrease towards the high-mass end beyond $\logMstarlimit \sim 10.4$. 
The decreasing evolution of satellite fractions at the massive end can be understood as a result of the short dynamical friction timescale of galaxies within massive dark haloes. 
The satellite fraction at $z>1$ is significantly reduced from $z\sim1$ even for less-massive galaxies, which is also seen in $z>3$ results using LBG samples. 
This may be due to the rarity of massive galaxies in the high-$z$; i.e., less-massive galaxies live as central galaxies rather than satellite galaxies in the vicinity of massive galaxies, which are quite rare objects compared to the low-$z$ Universe. 
Moreover, the stellar-mass loss and/or dimming of high-$z$ satellite galaxies due to the efficient tidal stripping/shocking, and the high merger rate can also be attributed to the small satellite fraction at $z>1$. 

\item The large-scale galaxy biases monotonically increase with the stellar-mass limit and redshift. 
The galaxy biases of the less-massive galaxies at $0.30 \leq z < 0.55$ show almost unity, indicating that low-mass galaxies in the low-$z$ Universe traces the underlying invisible dark matter distribution well. 

\item The stellar-to-halo mass ratio (SHMR) at $z \lesssim 1$ shows good agreement with the prediction of the empirical model by \citet{behroozi18}, although the results show a slight excess with $1.5-2\sigma$ levels at the massive ends. 
The discrepancy may be caused by an underestimate of the massive galaxy abundance in previous studies. 
SHMRs at a lower mass than the pivot halo mass show relatively smaller values compared to the theoretical predictions, although all of our results are within the $1\sigma$ confidence intervals of \citet{behroozi18}. 

\item Tracing the redshift evolution of the SHMRs reveals that the overall shape of the SHMR exhibits little evolution at $0<z<0.8$. 
However, we can clearly identify two characteristics beyond $z=0.8$: the peak SHMR decreases and the pivot halo mass shifts towards a higher halo mass with redshift. 
The redshift evolution of the SHMR can be explained by the redshift evolution of stellar-mass function and the supernova feedback effect. 
By connecting our SHMRs to higher-$z$ results, we systematically show that the pivot halo mass is almost constant $(\Mpivot \sim 10^{12}\Msun)$ up to $z\sim5$, using only the clustering and HOD-model analyses. 
This result is consistent with the theoretical predictions for galaxy formation models \citep[e.g.,][]{ree77,wang13,dekel19}. 

\item The instantaneous baryon conversion efficiencies (BCEs) of our SM samples are almost consistent with the predictions of empirical models. 
The halo mass with the peak BCE is $10^{11.0-11.5}h^{-1}\Msun$, which is $\sim 0.5-1.0$ dex smaller than the $\Mpivot$ of the SHMR. 
This indicates that the most efficient star-formation activity is thought to proceed in galaxies hosted by $10^{11.0-11.5}h^{-1}\Msun$ dark haloes, rather than galaxies that reside in $\Mpivot$ dark haloes. 

\end{enumerate}
Subsequent to this study, we will extend our research to less-massive galaxies in the same redshift range as well as higher-$z$ unbiased galaxies at $1.4<z<3.0$ using the dataset of the HSC SSP Deep/UltraDeep layers by combining with NIR data obtained by other extensive surveys (e.g., UKIDSS, VIKING) and $u$-band data obtained in the CLAUDS program \citep{sawicki19}. 
Moreover, we plan to investigate the relationship between dark haloes and galaxies more precisely by separating galaxy populations into star-forming galaxies and passive galaxies, using more sophisticated halo-model analysis methods \citep[e.g.,][]{,tinker13,cowley19}. 
By connecting the results in this study to future studies, we will be able to reveal the galaxy--dark halo co-evolution history across a wide mass range and cosmic time. 

\acknowledgments
We are grateful to Kohji Tomisaka, Yutaka Hayano, Yuichi Matsuda, Kazuhiro Shimasaku, Tsutomu T. Takeuchi, Masatoshi Imanishi, Takashi Hamana, Yuichi Harikane, and Masami Ouchi for discussions and giving us useful comments on this study. 

The Hyper Suprime-Cam (HSC) collaboration includes the astronomical communities of Japan and Taiwan, and Princeton University. 
The HSC instrumentation and software were developed by the National Astronomical Observatory of Japan (NAOJ), the Kavli Institute for the Physics and Mathematics of the Universe (Kavli IPMU), the University of Tokyo, the High Energy Accelerator Research Organization (KEK), the Academia Sinica Institute for Astronomy and Astrophysics in Taiwan (ASIAA), and Princeton University. 
Funding was contributed by the FIRST program from Japanese Cabinet Office, the Ministry of Education, Culture, Sports, Science and Technology (MEXT), the Japan Society for the Promotion of Science (JSPS), Japan Science and Technology Agency (JST), the Toray Science Foundation, NAOJ, Kavli IPMU, KEK, ASIAA, and Princeton University. 

This paper makes use of software developed for the Large Synoptic Survey Telescope. We thank the LSST Project for making their code available as free software at  http://dm.lsst.org

The Pan-STARRS1 Surveys (PS1) have been made possible through contributions of the Institute for Astronomy, the University of Hawaii, the Pan-STARRS Project Office, the Max-Planck Society and its participating institutes, the Max Planck Institute for Astronomy, Heidelberg and the Max Planck Institute for Extraterrestrial Physics, Garching, The Johns Hopkins University, Durham University, the University of Edinburgh, Queen's University Belfast, the Harvard-Smithsonian Center for Astrophysics, the Las Cumbres Observatory Global Telescope Network Incorporated, the National Central University of Taiwan, the Space Telescope Science Institute, the National Aeronautics and Space Administration under Grant No. NNX08AR22G issued through the Planetary Science Division of the NASA Science Mission Directorate, the National Science Foundation under Grant No. AST-1238877, the University of Maryland, and Eotvos Lorand University (ELTE) and the Los Alamos National Laboratory. 

This paper is based on data collected at the Subaru Telescope and retrieved from the HSC data archive system, which is operated by Subaru Telescope and Astronomy Data Center at National Astronomical Observatory of Japan. Data analysis was in part carried out with the cooperation of Center for Computational Astrophysics, National Astronomical Observatory of Japan. 

We acknowledge the Virgo Consortium for making their simulation data available. 
The EAGLE simulations were performed using the DiRAC-2 facility at Durham, managed by the ICC, and the PRACE facility Curie based in France at TGCC, CEA, Bruy\`{e}res-le-Ch\^{a}tel. 

SI acknowledges the support by JSPS KAKENHI Grant-in-Aid for Research Activity Start-up (17H07325). 

Numerical computations are in part carried out on the Cray XC30 (Aterui), the Cray XC50 (Aterui~II), and the GPU cluster operated by the Center for Computational Astrophysics (CfCA), National Astronomical Observatory of Japan. 
Data analyses are in part carried out on the open use data analysis computer system at the Astronomy Data Center, ADC, of the National Astronomical Observatory of Japan. 
This study has utilized the following open-source {\tt Python} packages: {\tt NumPy}, {\tt SciPy}, {\tt CosmoloPy}, {\tt CosmicPy}, {\tt AstroPy} \citep{astropy13,astropy18}, and {\tt hmf} \citep{murray13}. 

\facility{Subaru Telescope (Hyper Suprime-Cam)}.

\begin{longrotatetable}
\begin{deluxetable*}{llcccccccc}
\tablecaption{Best-fitting HOD Parameters of Cumulative Stellar-mass Limited Samples}
\tablehead{Redshift & Stellar-mass limit & $\log_{10}\Mmin$ & $\log_{10}M_{1}$ & $\log_{10}M_{0}$ & $\sigmaM$ & $\alpha$ & $\fsat$ & $\bg$ & $\chi^{2}$/d.o.f.}
\startdata
$0.30 \leq z < 0.55$ & $8.6$ & $11.128^{+0.083}_{-0.097}$ & $12.223^{+0.104}_{-0.088}$ & $12.444^{+0.162}_{-0.160}$ & $0.533^{+0.220}_{-0.388}$ & $0.735^{+0.467}_{-0.165}$ & $0.131 \pm 0.035$ & $1.090 \pm 0.033$ & $0.338$ \\
 & $8.8$ & $11.271^{+0.066}_{-0.063}$ & $12.808^{+0.063}_{-0.054}$ & $9.074^{+2.095}_{-3.116}$ & $0.472^{+0.412}_{-0.380}$ & $1.110^{+0.074}_{-0.181}$ & $0.164 \pm 0.029$ & $1.117 \pm 0.023$ & $0.823$ \\
 & $9.0$ & $11.381^{+0.050}_{-0.051}$ & $12.918^{+0.060}_{-0.049}$ & $7.885^{+2.603}_{-2.643}$ & $0.411^{+0.266}_{-0.264}$ & $1.192^{+0.217}_{-0.285}$ & $0.152 \pm 0.022$ & $1.141 \pm 0.029$ & $0.663$ \\
 & $9.2$ & $11.465^{+0.043}_{-0.038}$ & $12.946^{+0.065}_{-0.055}$ & $8.365^{+2.232}_{-2.307}$ & $0.249^{+0.182}_{-0.218}$ & $1.206^{+0.125}_{-0.498}$ & $0.171 \pm 0.017$ & $1.172 \pm 0.029$ & $0.497$ \\
 & $9.4$ & $11.542^{+0.046}_{-0.041}$ & $12.887^{+0.091}_{-0.069}$ & $9.694^{+1.988}_{-3.766}$ & $0.271^{+0.256}_{-0.172}$ & $1.090^{+0.111}_{-1.000}$ & $0.208 \pm 0.018$ & $1.176 \pm 0.040$ & $0.692$ \\
 & $9.6$ & $11.645^{+0.047}_{-0.042}$ & $13.108^{+0.078}_{-0.059}$ & $8.413^{+2.354}_{-2.365}$ & $0.262^{+0.173}_{-0.155}$ & $1.239^{+0.086}_{-0.078}$ & $0.159 \pm 0.017$ & $1.189 \pm 0.015$ & $0.373$ \\
 & $9.8$ & $11.738^{+0.044}_{-0.040}$ & $13.156^{+0.083}_{-0.064}$ & $8.303^{+2.174}_{-2.243}$ & $0.310^{+0.177}_{-0.191}$ & $1.218^{+0.074}_{-0.182}$ & $0.160 \pm 0.018$ & $1.196 \pm 0.018$ & $0.368$ \\
 & $10.0$ & $11.858^{+0.053}_{-0.044}$ & $13.260^{+0.109}_{-0.076}$ & $8.534^{+2.389}_{-2.462}$ & $0.253^{+0.180}_{-0.189}$ & $1.215^{+0.143}_{-0.387}$ & $0.158 \pm 0.021$ & $1.213 \pm 0.024$ & $0.202$ \\
 & $10.2$ & $12.035^{+0.055}_{-0.044}$ & $13.463^{+0.116}_{-0.092}$ & $8.628^{+2.513}_{-2.482}$ & $0.148^{+0.243}_{-0.108}$ & $1.238^{+0.172}_{-0.371}$ & $0.138 \pm 0.034$ & $1.241 \pm 0.051$ & $0.171$ \\
 & $10.4$ & $12.304^{+0.050}_{-0.042}$ & $13.683^{+0.069}_{-0.056}$ & $8.624^{+2.485}_{-2.465}$ & $0.235^{+0.205}_{-0.219}$ & $1.387^{+0.104}_{-0.375}$ & $0.119 \pm 0.031$ & $1.296 \pm 0.071$ & $0.150$ \\
 & $10.6$ & $12.624^{+0.048}_{-0.041}$ & $13.966^{+0.133}_{-0.066}$ & $9.078^{+2.756}_{-2.803}$ & $0.446^{+0.250}_{-0.361}$ & $1.500^{+0.162}_{-0.492}$ & $0.077 \pm 0.021$ & $1.324 \pm 0.043$ & $0.232$ \\
 & $10.8$ & $12.996^{+0.048}_{-0.039}$ & $14.480^{+0.135}_{-0.179}$ & $9.466^{+2.952}_{-3.054}$ & $0.486^{+0.266}_{-0.344}$ & $1.031^{+0.324}_{-0.316}$ & $0.047 \pm 0.028$ & $1.403 \pm 0.062$ & $0.204$ \\
 & $11.0$ & $13.506^{+0.046}_{-0.040}$ & $14.959^{+0.135}_{-0.509}$ & $10.168^{+2.991}_{-3.703}$ & $0.621^{+0.156}_{-0.523}$ & $1.161^{+0.222}_{-0.274}$ & $0.026 \pm 0.025$ & $1.541 \pm 0.127$ & $1.238$ \\
\hline
 $0.55 \leq z < 0.80$ & $9.0$ & $11.307^{+0.097}_{-0.067}$ & $12.235^{+0.203}_{-0.318}$ & $12.052^{+0.214}_{-0.368}$ & $0.233^{+0.112}_{-0.112}$ & $0.761^{+0.161}_{-0.158}$ & $0.201 \pm 0.038$ & $1.174 \pm 0.029$ & 0.139$$ \\
 & $9.2$ & $11.428^{+0.097}_{-0.072}$ & $12.445^{+0.142}_{-0.106}$ & $12.100^{+0.225}_{-0.308}$ & $0.235^{+0.108}_{-0.120}$ & $0.846^{+0.119}_{-0.149}$ & $0.190 \pm 0.036$ & $1.194 \pm 0.021$ & $0.075$ \\
 & $9.4$ & $11.538^{+0.093}_{-0.072}$ & $12.651^{+0.118}_{-0.092}$ & $11.990^{+0.638}_{-1.363}$ & $0.227^{+0.120}_{-0.119}$ & $0.959^{+0.186}_{-0.146}$ & $0.196 \pm 0.027$ & $1.216 \pm 0.026$ & $0.120$ \\
 & $9.6$ & $11.626^{+0.054}_{-0.042}$ & $12.891^{+0.075}_{-0.058}$ & $8.900^{+2.271}_{-2.772}$ & $0.270^{+0.008}_{-0.116}$ & $0.743^{+0.076}_{-0.087}$ & $0.211 \pm 0.016$ & $1.174 \pm 0.017$ & $0.332$ \\
 & $9.8$ & $11.711^{+0.046}_{-0.040}$ & $12.929^{+0.062}_{-0.052}$ & $10.869^{+1.685}_{-3.010}$ & $0.236^{+0.178}_{-0.153}$ & $1.134^{+0.101}_{-0.129}$ & $0.218 \pm 0.026$ & $1.245 \pm 0.018$ & $0.764$ \\
 & $10.0$ & $11.844^{+0.042}_{-0.038}$ & $12.993^{+0.063}_{-0.054}$ & $8.791^{+2.419}_{-2.616}$ & $0.365^{+0.159}_{-0.213}$ & $1.125^{+0.052}_{-0.072}$ & $0.221 \pm 0.022$ & $1.255 \pm 0.018$ & $0.320$ \\
 & $10.2$ & $12.041^{+0.046}_{-0.040}$ & $13.343^{+0.058}_{-0.048}$ & $8.568^{+2.453}_{-2.442}$ & $0.228^{+0.148}_{-0.202}$ & $1.287^{+0.186}_{-0.459}$ & $0.150 \pm 0.017$ & $1.294 \pm 0.023$ & $0.286$ \\
 & $10.4$ & $12.250^{+0.042}_{-0.038}$ & $13.478^{+0.075}_{-0.063}$ & $9.700^{+2.466}_{-3.054}$ & $0.278^{+0.249}_{-0.195}$ & $1.181^{+0.155}_{-0.582}$ & $0.155 \pm 0.032$ & $1.327 \pm 0.027$ & $0.810$ \\
 & $10.6$ & $12.565^{+0.047}_{-0.040}$ & $13.753^{+0.084}_{-0.064}$ & $8.751^{+2.551}_{-2.545}$ & $0.529^{+0.233}_{-0.413}$ & $1.318^{+0.175}_{-0.588}$ & $0.098 \pm 0.033$ & $1.349 \pm 0.055$ & $0.427$ \\
 & $10.8$ & $12.892^{+0.046}_{-0.039}$ & $14.231^{+0.231}_{-0.132}$ & $9.274^{+2.864}_{-2.945}$ & $0.597^{+0.172}_{-0.338}$ & $1.136^{+0.227}_{-0.355}$ & $0.054 \pm 0.025$ & $1.394 \pm 0.061$ & $0.455$ \\
 & $11.0$ & $13.311^{+0.060}_{-0.049}$ & $14.588^{+0.340}_{-0.361}$ & $9.845^{+3.179}_{-3.351}$ & $0.682^{+0.108}_{-0.346}$ & $1.136^{+0.233}_{-0.333}$ & $0.038 \pm 0.030$ & $1.498 \pm 0.088$ & $0.537$ \\
 & $11.2$ & $13.948^{+0.061}_{-0.046}$ & $15.377^{+0.278}_{-0.182}$ & $8.951^{+2.745}_{-2.653}$ & $0.840^{+0.104}_{-0.196}$ & $0.941^{+0.235}_{-0.174}$ & $0.017 \pm 0.009$ & $1.628 \pm 0.089$ & $0.696$ \\
\tablebreak
 $0.80 \leq z < 1.10$ & $9.4$ & $11.561^{+0.111}_{-0.080}$ & $12.779^{+0.141}_{-0.100}$ & $11.794^{+0.180}_{-0.177}$ & $0.330^{+0.187}_{-0.205}$ & $0.998^{+0.202}_{-0.323}$ & $0.148 \pm 0.028$ & $1.234 \pm 0.027$ & $0.191$ \\
 & $9.6$ & $11.691^{+0.065}_{-0.054}$ & $12.858^{+0.096}_{-0.077}$ & $11.949^{+0.151}_{-0.158}$ & $0.346^{+0.171}_{-0.218}$ & $1.002^{+0.216}_{-0.315}$ & $0.145 \pm 0.022$ & $1.258 \pm 0.026$ & $0.146$ \\
 & $9.8$ & $11.820^{+0.046}_{-0.030}$ & $13.184^{+0.058}_{-0.039}$ & $8.466^{+2.252}_{-2.315}$ & $0.116^{+0.064}_{-0.077}$ & $1.272^{+0.066}_{-0.113}$ & $0.133 \pm 0.010$ & $1.300 \pm 0.010$ & $0.171$ \\
 & $10.0$ & $11.931^{+0.042}_{-0.037}$ & $13.209^{+0.060}_{-0.050}$ & $8.137^{+2.044}_{-2.161}$ & $0.174^{+0.108}_{-0.145}$ & $1.267^{+0.047}_{-0.063}$ & $0.150 \pm 0.013$ & $1.329 \pm 0.012$ & $0.287$ \\
 & $10.2$ & $12.082^{+0.041}_{-0.036}$ & $13.283^{+0.058}_{-0.049}$ & $8.745^{+2.442}_{-2.556}$ & $0.267^{+0.087}_{-0.105}$ & $1.255^{+0.073}_{-1.000}$ & $0.156 \pm 0.028$ & $1.359 \pm 0.038$ & $0.163$ \\
 & $10.4$ & $12.332^{+0.041}_{-0.035}$ & $13.482^{+0.069}_{-0.062}$ & $8.942^{+2.567}_{-2.794}$ & $0.373^{+0.160}_{-0.196}$ & $1.253^{+0.112}_{-0.709}$ & $0.136 \pm 0.037$ & $1.403 \pm 0.029$ & $0.159$ \\
 & $10.6$ & $12.637^{+0.048}_{-0.039}$ & $13.776^{+0.104}_{-0.093}$ & $9.054^{+2.682}_{-2.869}$ & $0.522^{+0.231}_{-0.370}$ & $1.186^{+0.178}_{-0.373}$ & $0.097 \pm 0.031$ & $1.437 \pm 0.065$ & $0.391$ \\
 & $10.8$ & $12.963^{+0.047}_{-0.041}$ & $14.162^{+0.239}_{-0.161}$ & $9.496^{+2.842}_{-3.054}$ & $0.667^{+0.208}_{-0.366}$ & $1.096^{+0.243}_{-0.343}$ & $0.057 \pm 0.031$ & $1.465 \pm 0.092$ & $0.422$ \\
 & $11.0$ & $13.334^{+0.043}_{-0.035}$ & $14.657^{+0.233}_{-0.363}$ & $10.221^{+3.074}_{-3.417}$ & $0.667^{+0.228}_{-0.378}$ & $1.005^{+0.331}_{-0.326}$ & $0.038 \pm 0.032$ & $1.609 \pm 0.136$ & $0.382$ \\
 & $11.2$ & $13.713^{+0.041}_{-0.033}$ & $14.997^{+0.270}_{-0.185}$ & $9.537^{+3.058}_{-3.102}$ & $0.684^{+0.204}_{-0.361}$ & $0.987^{+0.355}_{-0.270}$ & $0.033 \pm 0.022$ & $1.786 \pm 0.185$ & $0.249$ \\ 
 \hline
$1.10 \leq z \leq 1.40$ & $9.8$ & $11.864^{+0.054}_{-0.052}$ & $13.340^{+0.043}_{-0.054}$ & $8.752^{+2.360}_{-2.587}$ & $0.102^{+0.020}_{-0.008}$ & $1.376^{+0.057}_{-0.166}$ & $0.074 \pm 0.008$ & $1.371 \pm 0.012$ & $0.624$ \\
 & $10.0$ & $11.979^{+0.075}_{-0.058}$ & $13.627^{+0.137}_{-0.126}$ & $8.740^{+2.492}_{-2.605}$ & $0.236^{+0.349}_{-0.155}$ & $1.017^{+0.165}_{-0.201}$ & $0.061 \pm 0.018$ & $1.369 \pm 0.030$ & $0.962$ \\
 & $10.2$ & $12.147^{+0.041}_{-0.041}$ & $13.493^{+0.128}_{-0.165}$ & $12.296^{+2.268}_{-3.921}$ & $0.096^{+0.114}_{-0.045}$ & $1.287^{+0.153}_{-0.275}$ & $0.061 \pm 0.023$ & $1.470 \pm 0.031$ & $0.129$ \\
 & $10.4$ & $12.483^{+0.044}_{-0.035}$ & $14.033^{+0.427}_{-0.270}$ & $5.058^{+2.801}_{-3.064}$ & $0.084^{+0.227}_{-0.043}$ & $1.441^{+0.245}_{-0.287}$ & $0.032 \pm 0.022$ & $1.598 \pm 0.045$ & $0.226$ \\
 & $10.6$ & $12.902^{+0.046}_{-0.046}$ & $14.475^{+0.294}_{-0.186}$ & $8.946^{+2.773}_{-2.636}$ & $0.576^{+0.091}_{-0.233}$ & $1.084^{+0.275}_{-0.255}$ & $0.024 \pm 0.014$ & $1.600 \pm 0.062$ & $0.380$ \\
 & $10.8$ & $13.287^{+0.040}_{-0.033}$ & $15.057^{+0.216}_{-0.160}$ & $8.563^{+2.706}_{-2.723}$ & $0.703^{+0.042}_{-0.126}$ & $0.883^{+0.168}_{-0.011}$ & $0.018 \pm 0.005$ & $1.676 \pm 0.046$ & $0.666$ \\
 & $11.0$ & $13.697^{+0.042}_{-0.037}$ & $14.979^{+0.289}_{-0.187}$ & $9.771^{+3.147}_{-3.252}$ & $0.847^{+0.097}_{-0.235}$ & $1.128^{+0.230}_{-0.266}$ & $0.010 \pm 0.008$ & $1.721 \pm 0.112$ & $0.263$ \\
 & $11.2$ & $13.989^{+0.040}_{-0.035}$ & $15.846^{+0.700}_{-1.778}$ & $10.933^{+2.477}_{-4.303}$ & $0.859^{+0.089}_{-0.184}$ & $0.780^{+0.149}_{-0.152}$ & $0.010 \pm 0.009$ & $1.847 \pm 0.112$ & $0.507$ \\ 
\enddata
\tablecomments{The stellar-mass limit is in units of $h^{-2}\Msun$ in a logarithmic scale, whereas all halo-mass parameters are in units of $h^{-1}\Msun$.}
\label{tab:Mstar_hod}
\end{deluxetable*}
\end{longrotatetable}

\startlongtable
\begin{deluxetable*}{llcccccc}
\tablecaption{Best-fitting HOD Parameters of Cumulative Star-formation Rate Limited Samples}
\tablehead{Redshift & SFR limit & $\log_{10}\Mmin$ & $\log_{10}M_{1}$ & $\log_{10}M_{0}$ & $\sigmaM$ & $\alpha$ & $\chi^{2}$/d.o.f.}
\startdata
$0.30 \leq z < 0.55$ & $-1.50$ & $10.720^{+0.194}_{-0.141}$ & $13.210^{+0.258}_{-0.640}$ & $9.553^{+2.135}_{-3.585}$ & $0.354^{+0.327}_{-0.196}$ & $0.843^{+0.230}_{-0.217}$ & $0.756$ \\
 & $-1.00$ & $10.954^{+0.273}_{-0.160}$ & $13.560^{+0.255}_{-0.406}$ & $8.805^{+2.489}_{-2.747}$ & $0.455^{+0.296}_{-0.203}$ & $0.784^{+0.208}_{-0.175}$ & $0.675$ \\
 & $-0.50$ & $11.122^{+0.102}_{-0.129}$ & $13.070^{+0.216}_{-0.244}$ & $9.880^{+1.831}_{-4.142}$ & $0.523^{+0.204}_{-0.340}$ & $0.959^{+0.212}_{-0.324}$ & $0.653$ \\
 & $0.00$ & $11.603^{+0.105}_{-0.081}$ & $13.596^{+0.263}_{-0.263}$ & $8.241^{+2.259}_{-2.237}$ & $0.468^{+0.277}_{-0.374}$ & $0.712^{+0.198}_{-0.114}$ & $0.799$ \\
 & $0.50$ & $12.440^{+0.165}_{-0.134}$ & $14.268^{+0.149}_{-0.203}$ & $9.148^{+2.885}_{-2.847}$ & $0.502^{+0.247}_{-0.339}$ & $1.146^{+0.227}_{-0.261}$ & $0.276$ \\
\hline
$0.55 \leq z < 0.80$ & $-1.50$ & $10.677^{+0.159}_{-0.120}$ & $11.936^{+0.131}_{-0.089}$ & $11.865^{+0.130}_{-0.136}$ & $0.480^{+0.244}_{-0.298}$ & $0.696^{+0.204}_{-0.125}$ & $0.298$ \\
 & $-1.00$ & $10.948^{+0.156}_{-0.105}$ & $12.185^{+0.176}_{-0.135}$ & $11.629^{+0.177}_{-0.223}$ & $0.507^{+0.223}_{-0.330}$ & $0.725^{+0.253}_{-0.148}$ & $0.476$ \\
 & $-0.50$ & $11.144^{+0.128}_{-0.105}$ & $12.380^{+0.208}_{-0.144}$ & $11.745^{+0.198}_{-0.446}$ & $0.481^{+0.249}_{-0.362}$ & $0.744^{+0.293}_{-0.167}$ & $0.159$ \\
 & $0.00$ & $11.535^{+0.066}_{-0.056}$ & $13.029^{+0.121}_{-0.106}$ & $8.499^{+2.288}_{-2.411}$ & $0.467^{+0.272}_{-0.276}$ & $0.993^{+0.178}_{-0.250}$ & $0.292$ \\
 & $0.50$ & $12.316^{+0.066}_{-0.054}$ & $14.288^{+0.428}_{-0.289}$ & $8.907^{+2.688}_{-2.580}$ & $0.574^{+0.262}_{-0.317}$ & $0.890^{+0.316}_{-0.256}$ & $0.154$ \\
\hline
$0.80 \leq z < 1.10$ & $-1.00$ & $11.051^{+0.095}_{-0.136}$ & $12.359^{+0.082}_{-0.093}$ & $10.999^{+1.843}_{-3.183}$ & $0.849^{+0.251}_{-0.346}$ & $0.946^{+0.033}_{-0.043}$ & $0.639$ \\
 & $-0.50$ & $11.299^{+0.091}_{-0.057}$ & $12.702^{+0.135}_{-0.079}$ & $7.482^{+1.729}_{-1.684}$ & $0.158^{+0.115}_{-0.119}$ & $1.112^{+0.058}_{-0.055}$ & $0.137$ \\
 & $0.00$ & $11.530^{+0.077}_{-0.068}$ & $12.915^{+0.117}_{-0.088}$ & $8.775^{+2.280}_{-2.699}$ & $0.512^{+0.226}_{-0.344}$ & $1.041^{+0.081}_{-0.290}$ & $0.220$ \\
 & $0.50$ & $12.086^{+0.055}_{-0.045}$ & $13.531^{+0.163}_{-0.118}$ & $8.565^{+2.401}_{-2.461}$ & $0.507^{+0.237}_{-0.348}$ & $1.038^{+0.217}_{-0.218}$ & $0.152$ \\
 & $1.00$ & $12.859^{+0.057}_{-0.047}$ & $14.315^{+0.113}_{-0.153}$ & $9.076^{+2.829}_{-2.823}$ & $0.523^{+0.222}_{-0.276}$ & $1.257^{+0.153}_{-0.223}$ & $0.103$ \\
\hline
$1.10 \leq z \leq 1.40$  & $-0.50$ & $11.383^{+0.058}_{-0.122}$ & $12.792^{+0.095}_{-0.115}$ & $8.160^{+2.126}_{-2.182}$ & $0.387^{+0.361}_{-0.263}$ & $1.131^{+0.081}_{-0.094}$ & $0.864$ \\
 & $0.00$ & $11.600^{+0.110}_{-0.070}$ & $13.085^{+0.113}_{-0.043}$ & $8.377^{+2.241}_{-2.342}$ & $0.143^{+0.129}_{-0.056}$ & $1.301^{+0.083}_{-0.139}$ & $0.304$ \\
 & $0.50$ & $12.059^{+0.061}_{-0.052}$ & $13.425^{+0.228}_{-0.155}$ & $9.691^{+2.111}_{-3.676}$ & $0.750^{+0.161}_{-0.695}$ & $1.069^{+0.227}_{-0.483}$ & $0.616$ \\
 & $1.00$ & $12.656^{+0.056}_{-0.049}$ & $14.050^{+0.280}_{-0.236}$ & $9.139^{+2.764}_{-2.807}$ & $0.824^{+0.136}_{-0.610}$ & $1.126^{+0.238}_{-0.295}$ & $0.341$ \\
\enddata
\tablecomments{The SFR limit is in units of $h^{-2}\Msun$yr$^{-1}$ in a logarithmic scale, whereas all halo-mass parameters are in units of $h^{-1}\Msun$.}
\label{tab:SFR_hod}
\end{deluxetable*}

\end{document}